\documentclass[sn-nature]{sn-jnl}

\usepackage{graphicx}%
\usepackage{floatrow}
\usepackage{bm}
\usepackage{multirow}%
\usepackage{subcaption}
\usepackage{siunitx}
\usepackage{amsmath,amssymb,amsfonts}%
\usepackage{amsthm}%
\usepackage{mathrsfs}%
\usepackage[title]{appendix}%
\usepackage[percent]{overpic}
\usepackage{xcolor}%
\usepackage{textcomp}%
\usepackage{manyfoot}%
\usepackage{booktabs}%
\usepackage{algorithm}%
\usepackage{algorithmicx}%
\usepackage{algpseudocode}%
\usepackage{listings}%
\usepackage{subcaption}
\usepackage{mathtools}
\usepackage{natbib}
\usepackage{multibib}

\begin{document}

\title[Article Title]{A Bayesian Framework to Investigate Radiation Reaction in Strong Fields}


\author*[1]{\fnm{E. E.} \sur{Los}} 

\author[2]{\fnm{C.} \sur{Arran}}
\author[1, 3, 4]{\fnm{E.} \sur{Gerstmayr}}
\author[3]{\fnm{M. J. V.} \sur{Streeter}}

\author[1]{\fnm{Z.} \sur{Najmudin}}%


\author[2]{\fnm{C. P.} \sur{Ridgers}}%

\author[3]{\fnm{G}. \sur{Sarri}}%
\author[1]{\fnm{S. P. D} \sur{Mangles}}%




\affil*[1]{\orgdiv{The John Adams Institute for  Accelerator Science}, \orgname{Imperial College  London}, \orgaddress{\street{Blackett Laboratory}, \city{London}, \postcode{SW72AZ}, \country{UK}}}

\affil[2]{\orgdiv{York Plasma Institute}, \orgname{University of York}, \orgaddress{\street{School of Physics, Engineering and Technology}, \city{York}, \postcode{YO10 5DD}, \country{UK}}}

\affil[3]{\orgdiv{School of Mathematics and Physics}, \orgname{Queen's University Belfast}, \orgaddress{\city{Belfast}, \postcode{BT7 1NN}, \country{UK}}}

\affil[4]{\orgdiv{Stanford PULSE Institute}, \orgname{SLAC National Accelerator Laboratory}, \orgaddress{\city{Menlo Park}, \postcode{CA 94025}, \country{USA}}}


%

%
\abstract{
Recent experiments aiming to measure phenomena predicted by strong field quantum electrodynamics have done so by colliding relativistic electron beams and high-power lasers.
In such experiments, measurements of the collision parameters are not always feasible, however, precise knowledge of these parameters is required for accurate tests of strong-field quantum electrodynamics.

Here, we present a novel Bayesian inference procedure which infers collision parameters that could not be measured on-shot. This procedure is applicable to all-optical non-linear Compton scattering experiments investigating radiation reaction. The framework allows multiple diagnostics to be combined self-consistently and facilitates the inclusion of prior or known information pertaining to the collision parameters. Using this Bayesian analysis, the relative validity of the classical, quantum-continuous and quantum-stochastic models of radiation reaction were compared for a series of test cases, which demonstrate the accuracy and model selection capability of the framework and and highlight its robustness in the event that the experimental values of fixed parameters differ from their values in the models. 
}









\maketitle

\section{Introduction}\label{sec:intro}


Laser-particle beam collisions have proven an effective tool in the investigation of strong-field quantum electrodynamics (SFQED). Early experiments utilised collisions between a linac-accelerated electron beam and a high-power laser~\cite{Burke_1997} to observe non-linear Compton scattering (NLCS)~\cite{Compton_1923} and the non-linear Breit-Wheeler (BW) process~\cite{Breit_1943}. More recently, so-called all-optical experiments (in which one laser pulse drives a wakefield accelerator~\cite{Tajima_1979}, producing a relativistic electron beam which collides with a second laser pulse) aimed to probe radiation reaction~\cite{poder_2018}\cite{cole_2018}, the recoil experienced by a charge accelerated in an external field upon emitting a photon. A number of initiatives~\cite{Yakimenko_2019}\cite{Abramowicz_2021} aim to perform high-precision studies of NLCS and non-linear BW pair creation.

The advent of laser facilities capable of attaining $\SI[]{e21}{\watt\per\cm\squared}-\SI[]{e23}{\watt\per\cm\squared}$ intensities~\cite{Willingale_2023,Gan_2021,Radier_2022,Papadopoulos_2019} will enable the exploration of SFQED in regimes where quantum effects are predicted to be substantial. Efforts are underway to measure vacuum birefringence in collisions between a high-power laser and a brilliant X-ray source~\cite{Karbstein_2022}\cite{Karbstein_2015}, while future campaigns propose to use high-power lasers to observe exotic phenomena as photon-photon splitting~\cite{Adler_1971} and the self-focussing and self-compression of light in vacuum~\cite{Marklund_2004a}. 

Many SFQED experiments propose to use collisions between high-power lasers and particles beams, or between high-power lasers, and may thus experience difficulties in interpreting data due to shot-to-shot variation in collision parameters. A recent study~\cite{Magnusson_2023} of the effect of varying collision parameters on model differentiation for radiation reaction studies using electron beam-laser collisions found $20\%-30\%$ changes in the mean and $5\%-10\%$ changes in the width of the final electron spectrum for typical variations in longitudinal alignment between the electron beam and laser pulse, while variations in the electron beam duration and chirp induced changes of up to $20\%$ in the final mean electron energy and width, respectively. This motivates the development of novel data analysis tools which account for uncertainties due to unknown or measured particle beam, laser and collision parameters self-consistently. 

To address this challenge, we have developed a Bayesian inference framework which facilitates parameter inference and model comparison for all-optical NLCS experiments aiming to probe radiation reaction. This framework infers values of collision parameters which directly affect experimental observables but were not measured on-shot, and incorporates knowledge of collision parameters from prior measurements or simulations (in the form of prior distributions). This procedure combines multiple diagnostics into a single, self-consistent analysis and enables a quantitative comparison of three radiation reaction models; the classical, quantum-continuous and quantum-stochastic models outlined in section~\ref{sec:theory}. While this framework applies to all-optical radiation reaction experiments, some of the techniques used in this work have wider relevance for beam-beam or laser-beam collider experiments.

Increasing the number of inference (or free) parameters rapidly increases the computational cost of the inference procedure beyond the point where the inference is tractable. Additionally, an excessive number of free parameters may result in overfitting. Therefore, a number of collision parameters are assigned fixed values. We assess the impact of fluctuating collision parameters on our experimental observables given their expected shot-to-shot variation. We then demonstrate that degeneracies between free and fixed parameters allow free parameters to replicate the effect on the experimental observables of a fixed parameter having an experimental value which differs from its value in the model. These two considerations inform the selection of free and fixed parameters.

When using this approach, inference parameters should be treated as effective parameters which replicate the collision conditions, rather than physical parameters which accurately represent electron beam and laser properties, or their spatio-temporal alignment.
If the experimental value of a fixed parameter deviates from its value in the model by more than a given amount, we find accurate model differentiation is no longer possible. We identify the threshold at which this occurs for transverse misalignments between the electron beam and colliding laser and propose to mitigate this issue by applying the Bayesian analysis to shots with the highest photon yields, for which transverse misalignments are likely to be small.
Finally, we assess the accuracy of model selection and parameter inference using the Bayesian framework via a variety of test cases with electron, laser and collision parameters representative of a recent experiment. We find that, given the experimental uncertainties and the broad, uniform priors we opted for, single-shot model differentiation is infeasible. However, model selection may be achieved by combining model evidences over multiple shots.

\section{Theory}
\label{sec:theory}

Two parameters, namely the electron quantum parameter, $\eta=E_{RF}/E_{sch}$, and the dimensionless intensity parameter, $a_0=\frac{|E_L|e}{\omega_L m_ec}$, govern the quantum and non-linear character of electron-photon interactions, respectively. The electron charge and mass are denoted $e$ and $m_e$, respectively, $\omega_L$ and $E_L$ are the frequency and strength of the external electric field in the lab frame respectively, $E_{RF}$ is the electric field strength in the electron rest frame and $E_{sch}=\SI[]{1.3e18}{\volt\per\metre}$ is the Schwinger field. A classical theory of radiation reaction is expected to be valid when $\alpha a_0\eta\simeq 1$ and $\eta \ll 1$~\cite{DiPiazza_2012}\cite{Blackburn_2020_b}, where $\alpha$ is the fine structure constant, while quantum effects become dominant for $\alpha a_0\simeq 1$ and $\eta\gtrsim1$~\cite{DiPiazza_2012}\cite{Blackburn_2020_b}. Multi-photon and relativistic effects manifest for $a_0>1$.

Classical radiation reaction is expected to be well-described by the Landau-Liftshitz model~\cite{LL_1971_ClassicalTheoryFields}, which treats radiation emission as a continuous process and does not impose an upper bound on the frequency of radiation emitted by an electron. For this reason, classical radiation reaction over-predicts electron energy loss compared to quantum models~\cite{Ridgers_2017}.

For $a_0>1$, quantum models of electron-photon scattering become non-perturbative. An alternative approach is employed (the Furry picture) in which electrons are ``dressed'' by the background field~\cite{Furry_1951}. In order to treat arbitrary electromagnetic fields, photon emission is assumed is occur over timescales much smaller than the timeframe of electromagnetic field variation, allowing the electric and magnetic fields to be treated as locally constant and crossed (locally-constant field approximation, or LCFA)~\cite{Ritus_1985}. Between photon emissions, electrons propagate classically~\cite{Duclous_2010}\cite{Arber_2015}\cite{Kirk_2009}. The quantum model of radiation reaction, here termed the quantum-stochastic model, prohibits the emission of photons with energies exceeding the electron energy and treats photon emission as a stochastic process, which gives rise to spectral broadening~\cite{Ridgers_2017}\cite{Niel_2018}\cite{Neitz_2013}. 

We also consider a quantum-continuous model, which incorporates first-order quantum effects in a classical framework. This model treats radiation emission as continuous but applies a correction factor, the Gaunt factor~\cite{Baier_1998}, to the radiation reaction force term to recover the same rate of change of average electron momentum predicted by the quantum-continuous model~\cite{Ridgers_2017}\cite{Niel_2018}. Both the classical and quantum-continuous models predict spectral narrowing~\cite{Ridgers_2017}\cite{Niel_2018}.

Throughout this paper, the subscripts $\rm{qs}$, $\rm{qc}$ and $\rm{cl}$ refer to the quantum-stochastic, quantum-continuous and classical models.

\section{Method}\label{sup:BI_details}

\subsection{Bayesian Statistics}

Bayesian inference is a statistical technique which allows the unknown variables, $\rho$, which parameterise a given model, $M^x$, to be inferred. Central to this technique is the approach of using a forward model (which predicts the experimental observables for a set of inputs) to obtain the probability that the model is accurate given the observed data, $D$, i.e. $P(M^x\mid D)$, the posterior probability. Crucially, the forward models, and hence their posterior probability distributions, are functions of the inference parameters. 

The posterior probability is calculated using Bayes' theorem:
\begin{equation}
P(M^x\mid D)=\frac{P(D\mid M^x)P(M^x)}{P(D)}
\end{equation}
where $P(D\mid M^x)$ is the likelihood of observing the data given $M^x$. The prior probability, $P(M^x)$, incorporates known information about the inference parameters, and the probability of observing the data, $P(D)$, is a normalisation constant.
Two models, here demarcated by $a$ and $b$, may be compared quantitatively using a Bayes factor, $r$, defined as
\begin{equation}
\label{eqn:Bayes_factor_define}
r_{a,b}=\frac{\int P(D\mid \rho_a, M^a)P(\rho_a\mid M^a)d\rho_a}{\int P(D\mid \rho_b, M^b)P(\rho_b\mid M^b)d\rho_b}
\end{equation}
Table~\ref{Bayes_factor_interpretation} provides guidelines for the interpretation of Bayes factors. Note that for $r_{a,b}>1$, $M_a$ is favoured over $M_b$, while for $r_{a,b}<1$, the reverse is true and the Bayes factor interpretation is given by the reciprocal of the first column in table~\ref{Bayes_factor_interpretation}.
\begin{table}[H]
\begin{tabular}{|c|c|}
\hline
Bayes factor & interpretation of result \\
\hline
1-3.2 & inconclusive \\
3.2-10& substantial\\
10-100& strong\\
$>$100& decisive\\
\hline
\end{tabular}
\caption[Bayes factor interpretation]{Guidelines for Bayes factor interpretation~\cite{Kass_1995}.}
\label{Bayes_factor_interpretation}
\end{table}

The Bayes factor may be challenging to compute as it requires integrals over $f$-dimensional space, where $f$ is the number of fit parameters. For this reason, $r$ is often approximated numerically. We used leave-one-out cross-validation with Pareto-smoothed importance sampling~\cite{Vehtari_2017}, available in the python package arviz~\cite{arviz_2019}, to compute Bayes factors.


A Markov Chain Monte Carlo (MCMC) algorithm, implemented using the emcee library in python~\cite{Foreman-Mackey_2013}, was used to perform the inference.

\subsection{Implementation of Bayesian Inference}\label{Implementation_Bayesian_inference}

\begin{figure}[ht!]%
\centering
\includegraphics[width=1.0\textwidth]{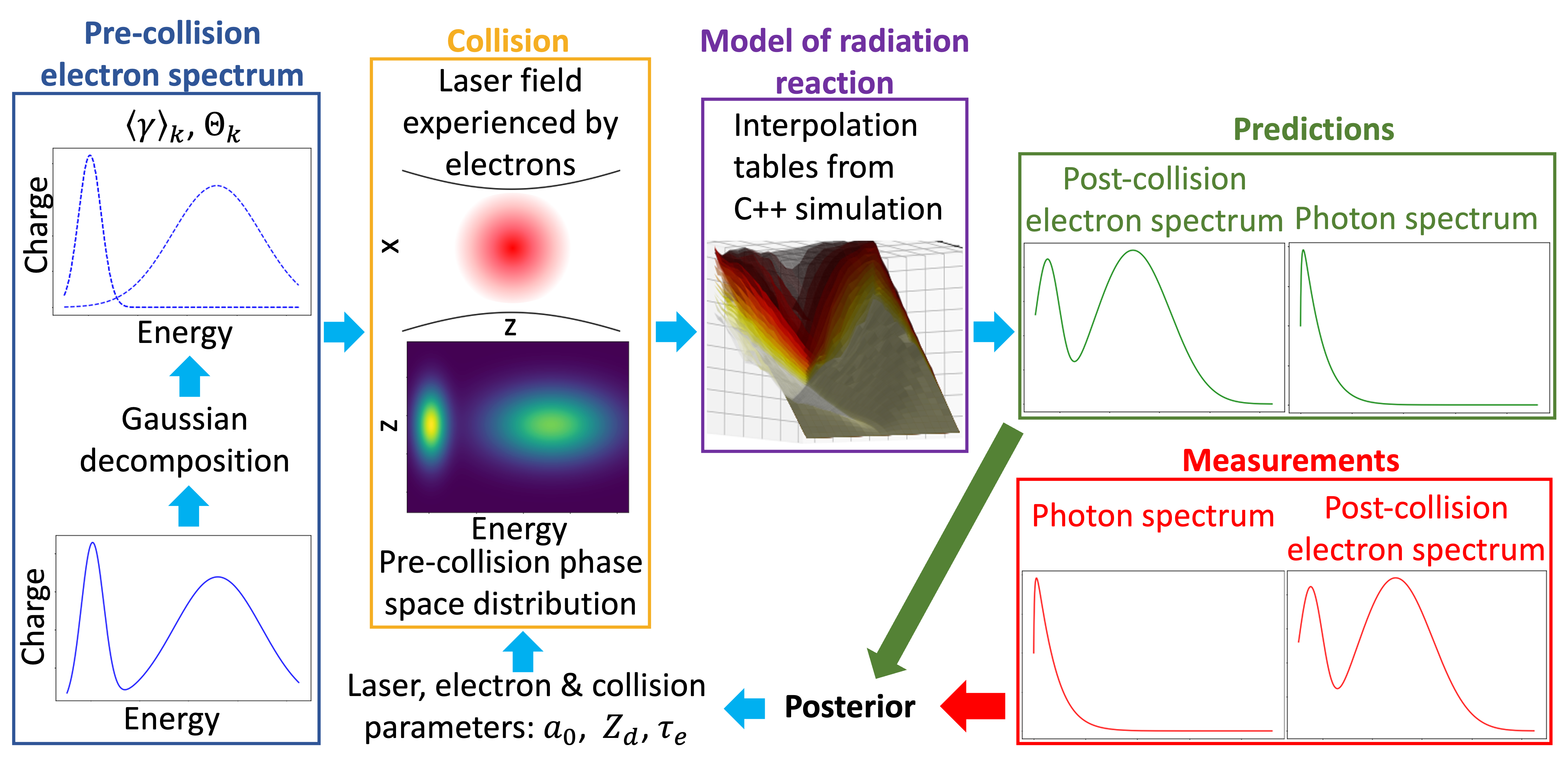}
\caption{The stages of the Bayesian analysis procedure are summarised. Initially, a distribution of pre-collision electron spectra are predicted by a neural network (for simplicity only one pre-collision spectrum is shown). The pre-collision spectrum is decomposed into a sum of Gaussian sub-bunches which are fed into the inference procedure. The MCMC returns three inference parameters; the laser $a_0$, longitudinal displacement of the collision from the laser focus, $Z_d$, and the electron beam duration, $\tau_e$, which are used to re-construct the pre-collision phase space of the electron beam and the laser electric field it experiences at the collision. This information is supplied to the forward model, which predicts the post-collision electron spectrum and photon spectrum for each sub-bunch. The full post-collision electron and photon spectra are obtained by performing a charge-weighted sum over the sub-spectra predicted for each sub-Gaussian. The model predictions, measured post-collision electron and photon spectra and their uncertainties are used to compute the posterior probability, which allows the MCMC algorithm to predict the subsequent region of the posterior to sample.\label{fig:flow_chart_num_methods}
}
\end{figure}

The Bayesian inference procedure is summarised in figure~\ref{fig:flow_chart_num_methods}. It was not possible to measure the pre-collision electron spectrum for successful collision shots. Instead, electron spectra were measured for shots where the colliding laser was not fired. These spectra were used to train a neural network, which was then used to predict pre-collision electron spectra for successful collision shots. The construction, training and testing of this neural network are discussed by Streeter et. al.~\cite{Streeter_2023}. The neural network consisted of an encoder followed by a translator stage and subsequently a decoder. The former compressed information from four diagnostics (the plasma density, laser energy and pointing and the recombination light emitted by the plasma) into the minimum number of parameters which allowed the key features of the inputs to be reconstructed. The decoder performed a similar function but in reverse, reconstructing a full pre-collision electron spectrum from a small number of inputs. The translator section provided a mapping between the outputs and inputs of the encoder and decoder, respectively. Once trained, the encoder, translator and decoder were used to predict an ensemble of pre-collision electron spectra. This was used to estimate the uncertainty due to the availability of training data for each collision shot analysed using the Bayesian inference procedure. The near-median and standard deviation of the predicted distribution (the former is defined as the spectrum closest in shape to the median spectrum of the distribution) were used to approximate the pre-collision electron spectrum and its uncertainty.

In the following section, curled variables, such as $\mathcal{N}$, denote post-collision variables and pre-collision variables are italicised. The superscripts $\bar{}$ and $'$ demarcate properties of the gamma spectrum and measured (as opposed to predicted) observables, respectively.

The pre-collision electron spectra were complex and varied significantly from shot-to-shot, necessitating an analysis procedure capable of treating arbitrary electron spectra. To this end, a routine was developed which decomposed pre-collision electron spectra into a sum over Gaussian sub-spectra, with mean and standard deviation Lorentz factors, $\langle\gamma\rangle_k$ and $\Theta_k$, respectively, where the subscript $k$ iterates over the number of sub-spectra, $n_{\langle \gamma \rangle}$. The sub-spectra were fed into the MCMC algorithm which sampled three inference parameters from their prior distributions, namely the laser $a_0$, the longitudinal offset of the collision from the laser focus, $Z_d$, and the electron bunch duration, $\tau_e$. The Gaussian sub-spectra and $\tau_e$ were used to obtain the electron beam phase-space distribution. The spatio-temporal distribution of the laser intensity at the collision was derived using $a_0$ and $Z_d$, assuming gaussian spatial and temporal profiles and a Gouy phase term for the laser, and no transverse offset of the collision from the laser focus. The laser $a_0$ and electron beam phase-space distribution comprised the inputs for the forward model, i.e. the parameterised model of radiation reaction. 

Each forward model consisted of five 4-dimensional interpolation tables produced using a Monte Carlo code written in C++ (see supplementary note~\ref{sup:RR_models}). Three of these tables parameterise the post-collision electron spectrum as a Weibull distribution (as indicated by simulations), providing its location, $\mu$, scale, $\lambda$, and shape, $\kappa$,
\begin{equation}
\label{eqn:weibul}
\frac{\rm{d}\mathcal{N}}{\rm{d}\gamma}=\frac{\kappa}{\lambda}\left(\frac{\gamma-\mu}{\lambda}\right)^{\kappa-1}e^{-\left(\frac{\gamma-\mu}{\lambda}\right)^{\kappa}}.
\end{equation}

The two remaining interpolation tables returned the photon number, $A$, and the normalised critical energy, $\bar{\epsilon}_{c}$, of the photon spectrum,
\begin{equation}
\label{eqn:gamma_spec}
\frac{\rm{d}\bar{\mathcal{F}}_{k, l}}{\rm{d}{\bar{\epsilon}}}=A\left(\frac{\bar{\epsilon}}{\bar{\epsilon}_c}\right)^{-\frac{2}{3}}e^{-\frac{\bar{\epsilon}}{\bar{\epsilon}_c}},
\end{equation}
where $\bar{\epsilon}=\frac{\hbar\omega}{m_ec^2}$.

\begin{figure}[ht!]%
\centering
\begin{overpic}[width=1.0\textwidth, trim={0.8cm 0.0cm 0.45cm 0.6cm}, clip]{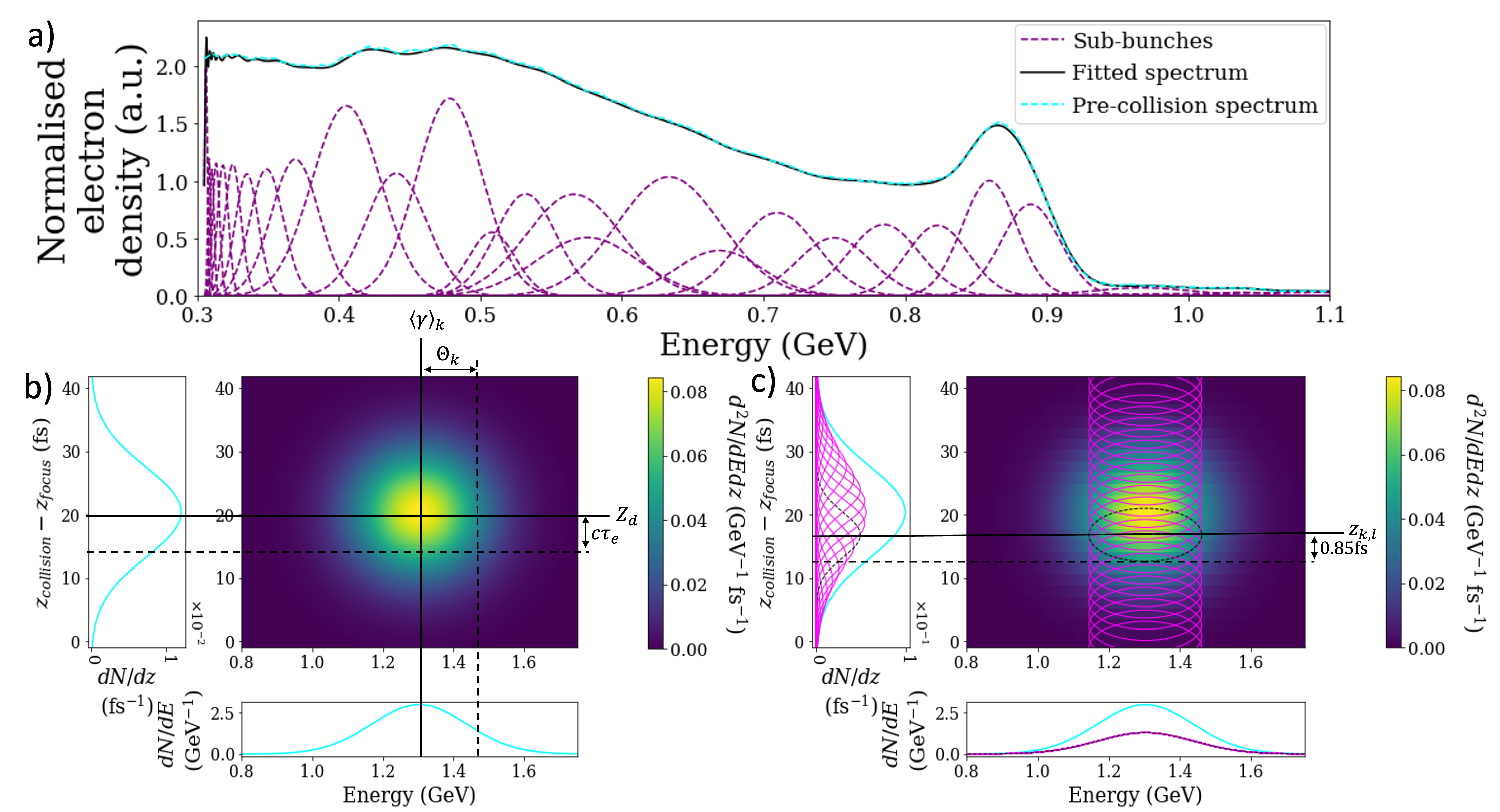}
\caption{a) The decomposition of a pre-collision electron spectrum predicted by a neural network (cyan) into Gaussian sub-spectra (purple), the sum over which (black) reproduces the original spectrum. b) The phase space projection (centre) of a single gaussian sub-spectrum with mean, $\langle\gamma\rangle_k$, and standard deviation, $\Theta_k$, Lorentz factor demarcated by continuous and dashed vertical black lines, respectively. The location, $Z_d$, and width, $c\tau_e$, of its longitudinal distribution are indicated by continuous and dashed horizontal black lines, respectively. The longitudinal (left) and spectral (bottom) distributions of the gaussian sub-spectrum (obtained by integrating its phase-space distribution over the spectral and longitudinal axes, respectively) are show in cyan. c) Decomposition of the phase-space distribution in b) into femto-bunches (magenta) with varying numbers of electrons, $g_{k, l}$, evenly-spaced mean longitudinal positions, $z_{k, l}$ and \SI{0.85}{\femto\second} durations, where the latter two properties are indicated for a single femto-bunch by continuous and dashed black horizontal lines. The sum over the femto-bunches yields the spectral (bottom) and longitudinal (left) distributions shown in cyan.
\label{fig:overview_phase_space_decomp}
}
\end{overpic}
\end{figure}

Pre-collision electron spectra input into the Bayesian inference procedure were decomposed into $n_{\langle \gamma \rangle}$ Gaussian sub-spectra which re-created the full spectra when summed, as illustrated in figure~\ref{fig:overview_phase_space_decomp}a):
\begin{equation}
\frac{\rm{d}N}{\rm{d}\gamma}=
\sum^{n_{\langle \gamma \rangle}}_{k=1}h_k e^{-\frac{(\gamma-\langle \gamma \rangle_k)^2}{2\Theta^2_k}}
\end{equation}
where $\langle \gamma \rangle_k$, $\Theta_k$ and $h_k$ denote the mean, standard deviation and electron number of the $k^{th}$ bunch, where k denotes decomposition along spectral axis, and $0\leq k\leq n_{\langle \gamma \rangle}$.

Given the longitudinal displacement of the collision from the laser focus, $Z_d$, and the electron bunch duration, $\tau_e$, sampled by the Bayesian inference procedure, the phase space of the pre-collision electron spectrum, obtained under the assumption that each sub-spectrum has a Gaussian longitudinal distribution as indicated in figure~\ref{fig:overview_phase_space_decomp}b), is given by:
\begin{equation}
\frac{\rm{d}^2N}{d\gamma \rm{d}z}=\sum^{n_{\langle \gamma \rangle}}_{k=1}\frac{h_k}{\sqrt{2\pi c^2\tau_e^2}}e^{-\frac{(\gamma-\langle \gamma \rangle_k)^2}{2{\Theta^2_k}}}e^{-\frac{\left(Z-Z_d\right)^2}{2c^2\tau_e^2}}
\end{equation}
where $Z$ denotes the longitudinal co-ordinate. The interpolation tables were produced using electron beams with gaussian temporal profiles and standard deviation durations $2/\omega_L=\SI[]{0.85}{\femto\second}$. To enable the inference procedure to construct electron beams of arbitrary duration without requiring the addition of a fifth dimensional interpolation table for $\tau_e$, a second Gaussian decomposition was performed on the electron spectrum input into the Bayesian analysis, in which each sub-bunch was split into $n_z$ femto-bunches of \SI[]{0.85}{\femto\second} duration, as shown in figure~\ref{fig:overview_phase_space_decomp}c). This second, longitudinal decomposition is denoted by the index $l$. The longitudinal position of each femto-bunch, $Z_d-3c\tau_e\leq z_{k, l}\leq Z_d+3c\tau_e$. 
The electron number in each femto-bunch, $g_{k, l}$, is given by:
\begin{equation}
g_{k, l}=\frac{h_k}{\sqrt{2\pi c\tau_e}}e^{-\frac{\left(\gamma-\langle \gamma \rangle_k\right)^2}{2\Theta^2_k}}e^{-\frac{\left(z_{k, l}-Z_d\right)^2}{2c^2\tau_e^2}}.
\end{equation}
Together, $a_0$, $g_{k, l}$, $\Theta_k$ and $\langle\gamma\rangle_k$ comprise the inputs for the interpolation tables which constitute the forward model. 

\begin{figure}[ht!]%
\centering
\begin{overpic}[width=1.0\textwidth, trim={0.4cm 0.15cm 0.6cm 0.55cm}, clip]{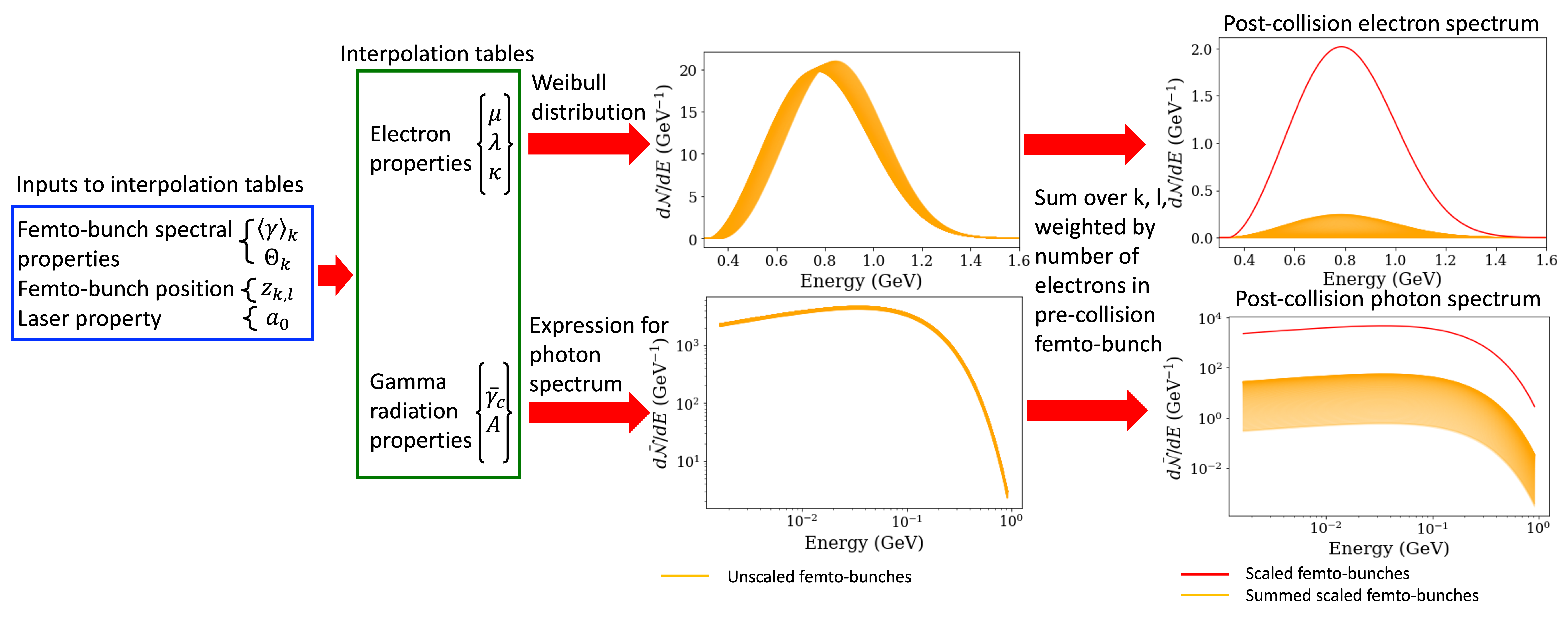}
\caption{Overview of the forward models used to predict the post-collision electron and photon spectra. Once the phase space decomposition has been performed, the mean and standard deviation Lorentz factor and mean longitudinal position  of each femto-bunch are fed into five interpolation tables together with the laser $a_0$. Each interpolation table generates a single output, three of which describe the post-collision electron spectrum location, $\mu$, scale, $\lambda$, and shape factor, $\kappa$, while the remaining tables output the critical factor, $\bar{\epsilon}_c$, and photon number, $A$, of the photon spectrum. The interpolation table outputs are used to obtain the post-collision electron and gamma spectra for each femto-bunch, which are then weighted by the number of electrons in the pre-collision femto-bunch and summed, yielding the full post-collision electron and photon spectra, respectively.\label{fig:overview_forward_model}
}
\end{overpic}
\end{figure}

The forward model for the post-collision electron spectrum consists of three interpolation tables which each generate one output for the four inputs which parameterise each femto-bunch. The three outputs, namely $\mu$, $\lambda$ and $\kappa$, are fed into equation~\ref{eqn:weibul} to obtain each post-collision femto-spectrum, $\frac{\rm{d}\mathcal{F}_{k, l}}{\rm{d}\gamma}$. This procedure is illustrated in figure~\ref{fig:overview_forward_model}.

As the interpolation tables returned normalised electron spectra, the post-collision electron femto-bunches are weighted by the number of electrons in the corresponding pre-collision femto-bunch and then summed over to obtain the full post-collision electron spectrum, $\frac{\rm{d}\mathcal{N}}{\rm{d}\gamma}$,
\begin{equation}
\label{eqn:charge_on_rescale}
\frac{\rm{d}\mathcal{N}}{\rm{d}\gamma}=\frac{1}{D_e} \sum_{k=1}^{n_{\langle \gamma \rangle}} \sum_{l=1}^{n_z}g_{k, l}\Theta_k\frac{\rm{d}\mathcal{F}_{k, l}}{\rm{d}\gamma}.
\end{equation}
where $D_e=\left(\sum_{k=1}^{n_z}g_{k, l}\Theta_k\right)$. The post-collision photon spectrum was obtained using a similar procedure, where the forward model returns two parameters per femto-bunch, $A$ and $\bar{\epsilon}_c$, which are inserted into equation~\ref{eqn:gamma_spec} to obtain $\frac{\rm{d}\bar{\mathcal{F}}_{k, l}}{\rm{d}{\bar{\epsilon}}}$. As with the electron spectrum, a weighted sum is performed to obtain the full photon spectrum, $\frac{\rm{d}\bar{\mathcal{N}}}{\rm{d}\bar{\epsilon}}$.

\subsection{Error propagation}

The neural network which predicted the pre-collision electron spectra returned $n_P=100$ distinct spectra, $\frac{\rm{d}N_j}{\rm{d}\gamma}$, where $j$ enumerates each prediction. At each iteration of the Bayesian inference procedure, this pre-collision distribution is used to obtain the corresponding post-collision electron spectra for a set of collision parameters as discussed in section~\ref{Implementation_Bayesian_inference}. The near-median of the distribution of post-collision spectra, $\frac{\rm{d}\mathcal{N}_{\text{nm}}}{\rm{d}\gamma}$, approximates the post-collision spectrum. The uncertainty on the predicted post-collision electron spectrum, $\zeta(\gamma)$, is the root mean squared (rms) deviation of the distribution of predicted post-collision spectra from the near-median spectrum.
The uncertainty in the measured post-collision electron spectrum due to the uncertainties in the magnet, lanex screens and gas jet positions, $\zeta'(\gamma)$ is
\begin{equation}
\label{eqn:electron_sigma_meas}
\zeta'(\gamma)=\frac{\left(\frac{\rm{d}\mathcal{N}_{\text{nm}}}{\rm{d}\gamma}(\gamma+\sigma_{\gamma})+\frac{\rm{d}\mathcal{N}_{\text{nm}}}{\rm{d}\gamma}(\gamma-\sigma_{\gamma})\right)}{2}
\end{equation}
where $\sigma_{\gamma}=C_e(\gamma m_e[\SI[]{}{\mega\electronvolt}])^2/m_e[\SI[]{}{\mega\electronvolt}]$, and $C_e=\SI[]{32.45e-6}{\per\mega\electronvolt}$. 
The log likelihood for the electron spectrum is then
\begin{equation}
\begin{aligned}
&\left\langle\log\left(P\left(\frac{\rm{d}\mathcal{N}'}{\rm{d}\gamma} \mathrel{\Big|} M^{x}\right)\right)\right\rangle= \\
&\int  \left(-\frac{1}{2}\log\left(2\pi\left[\zeta^2(\gamma)+\zeta'^2(\gamma)\right]\right)-\frac{\left(\frac{\rm{d}\mathcal{N}'}{\rm{d}\gamma} - \frac{\rm{d}\mathcal{N}_{\text{nm}}}{\rm{d}\gamma}\right)^2}{2\left(\zeta^2(\gamma)+\zeta'^2(\gamma)\right)}\right)d\gamma
\end{aligned}
\end{equation}
where $\frac{\rm{d}\mathcal{N}'}{\rm{d}\gamma}$ is the measured post-collision electron spectrum.

The uncertainty, $\bar{\zeta}_{pre}$, in the predicted post-collision gamma spectrum, $\frac{\rm{d}\bar{\mathcal{N}}}{\rm{d}\bar{\epsilon}}$, due to the uncertainty in the predicted pre-collision electron spectrum is given by the standard deviation of the distribution of predicted post-collision gamma spectra.

A separate Bayesian inference routine was used to fit equation~\ref{eqn:gamma_spec} given the measured gamma spectrometer signal. This procedure yielded $n_\mathcal{Y}$-valued distributions of $\bar{\epsilon}_c$ and $A$ which yielded the best fit. 
These values of $\bar{\epsilon}_c$ and $A$ were used to generate a distribution of $n_\mathcal{Y}$ gamma spectra, $\frac{\rm{d}\mathcal{N}'_{v}}{\rm{d}\bar{\epsilon}}$, where $v$, which denotes the $v^{th}$ spectrum in the distribution, has values $0\leq v\leq n_\mathcal{Y}$. The mean, $\left\langle\frac{\rm{d}\bar{\mathcal{N}}'}{\rm{d}\bar{\epsilon}}\right\rangle$, and standard deviation, $\bar{\zeta}'(\bar{\epsilon})$, of the distribution indicate the most probable gamma spectrum and its corresponding uncertainty.
The log likelihood, $\left\langle\log\left(p\left(\frac{\rm{d}\bar{\mathcal{N}}'}{\rm{d}\bar{\epsilon}}\mathrel{\Big|}M^{x}\right)\right)\right\rangle$, for the gamma spectrum is then
\begin{equation}
\label{eqn:LL_gs}
\begin{aligned}
&\left\langle\log\left(p\left(\frac{\rm{d}\bar{\mathcal{N}}'}{\rm{d}\bar{\epsilon}}\mathrel{\Big|}M^{x}\right)\right)\right\rangle \\
&=\frac{1}{n_P}\sum_{j=0}^{n_P}\frac{1}{\Delta\bar{\epsilon}}\int \left[-0.5\log\left(2\pi \bar{\zeta}^2\left(\bar{\epsilon}\right)\right)-\frac{\left(\frac{\rm{d}\bar{\mathcal{N}}_j}{\rm{d}\bar{\epsilon}}-\frac{\rm{d}\bar{\mathcal{N}}'}{\rm{d}\bar{\epsilon}}\right)^2}{2\bar{\zeta}^2\left(\bar{\epsilon}\right)}\right] d\bar{\epsilon}
\end{aligned}
\end{equation}
where $\bar{\zeta}\left(\bar{\epsilon}\right)=\sqrt{\bar{\zeta}^2_{\text{pre}}\left(\bar{\epsilon}\right)+\bar{\zeta}'^2\left(\bar{\epsilon}\right)}$, and $\Delta\bar{\epsilon}$ is the interval size for $\bar{\epsilon}$. The total likelihood, $\log(P(D_{T}\mid M^{x}))$, is then
\begin{equation}
\begin{aligned}
&\log(P(D_{T}\mid M^{x}))=\frac{1}{2}\left\langle\log\left(P\left(\frac{\rm{d}\bar{\mathcal{N}}'}{\rm{d}\bar{\epsilon}}\mathrel{\Big|}M^{x}\right)\right)\right\rangle\\
&+\frac{1}{2}\left\langle\log\left(P\left(\frac{\rm{d}\mathcal{N}'}{\rm{d}\gamma}\mathrel{\Big|}M^{x}\right)\right)\right\rangle.
\end{aligned}
\end{equation}
where $D_{T}$ denotes the measured electron and gamma spectra.

The prior distribution, $P(a_0, Z_d, \tau_e)$, is given by the sum over the log priors for $a_0$, $P(a_0)$ and $z_{k, l}$, $P(z_{k, l})$:
\begin{equation}
\log(P(a_0, Z_d, \tau_e))=\log(P(a_0))+\log(P(z_{k, l}))
\end{equation}
where $P(a_0)$, $P(z_{k, l})$ are uniform priors which reflect the ranges of the interpolation tables for $a_0$ and $z_{k, l}$, respectively:
\begin{align}
\log(P(a_0))&=0 \quad 0.1\leq a_0 \leq 35\nonumber\\
\log(P(a_0))&=-\infty \quad a_0 < 0.1\\
\log(P(a_0))&=-\infty \quad a_0 > 35\nonumber\\
\noalign{ }
\log(P(z_{k, l}))&=0 \quad -4800 /\omega_L\leq z_{k, l} \leq 1440/\omega_L\nonumber\\
\log(P(z_{k, l}))&=-\infty \quad z_{k, l}\leq -4800 /\omega_L\\
\log(P(z_{k, l}))&=-\infty \quad z_{k, l} \geq 1440/\omega_L.\nonumber\
\end{align}

The quantity optimised by the MCMC, $\log(P\left(M^{x}\mid D_{T})\right)$, which is proportional to the posterior probability, was calculated using
\begin{equation}
\log(P\left(M^{x}\mid D_{T})\right))=\log(P(D_{T}\mid M^{x}))+\log(P(a_0, z_{k, l})).
\end{equation}
The MCMC optimised the log posterior as this enabled the inference procedure to treat extremely small probabilities. The $\log(P\left(M^{x}\mid D_{T})\right)$ obtained for a given set of $a_0$, $Z_d$ and $\tau_e$ determined the subsequent region of the parameter space to be explored by the MCMC. This procedure was continued iteratively until the MCMC converged. The degree of convergence and the point at which convergence was reached were calculated as described by Sokal et al.~\cite{Sokal_1996} and the emcee module documentation~\cite{emcee_autocorrelation}, respectively. 

\subsection{Benchmarks for Forward Models used by Bayesian Analysis}

\begin{figure}[ht!]
\centering
\begin{floatrow}
\ffigbox[\FBwidth]{\caption{The post-collision electron spectrum obtained from a Monte Carlo simulation for a collision between an electron beam with initial $\langle\gamma\rangle=2550$ and $\Theta_k=263.4$ and a laser with $a_0=35$ where $Z_d=0$. The reconstructed electron spectrum obtained using the interpolation tables (magenta) shows good agreement with the simulated post-collision spectrum.\label{fig:espec_recon_benchmark_RR2021}}{\includegraphics[width=6.5cm, height=5cm, trim={0.3cm 0.3cm 0.3cm 0.3cm},clip]{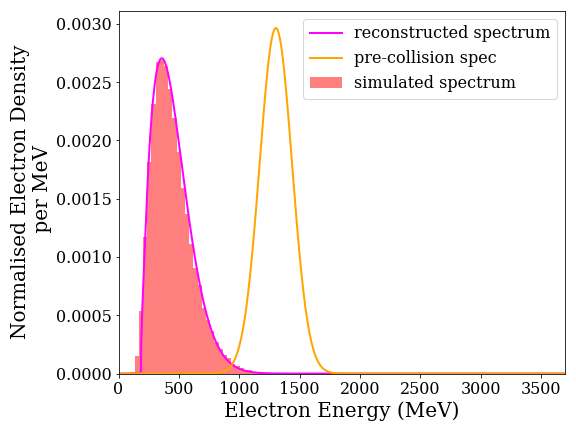}}}

\ffigbox[\FBwidth]{\caption{The photon spectrum simulated using a Monte Carlo code for the parameters provided in figure~\ref{fig:espec_recon_benchmark_RR2021} is shown alongside the fit thereto (with equation~\ref{eqn:gamma_spec}) and the photon spectrum constructed using the interpolation tables.\label{fig:gspec_recon_benchmark_RR2021}}{\includegraphics[width=6.5cm, height=5cm, trim={0.1cm 0.2cm 1.2cm 1.2cm},clip]{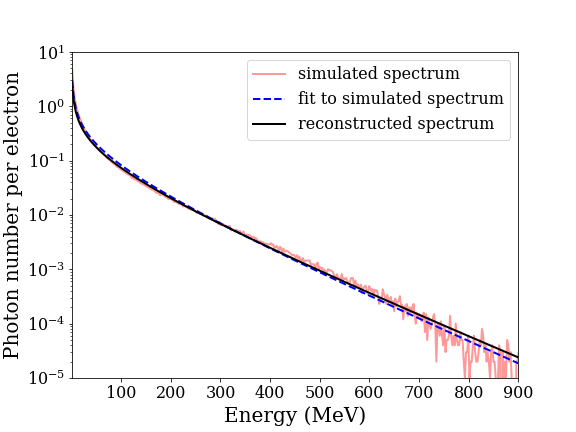}}}

\end{floatrow}
\end{figure}

The full forward models were benchmarked using the Monte Carlo code QEDCASCADE~\cite{QEDCASCADE_2021}~\cite{Watt_2021}, as illustrated in figures~\ref{fig:espec_recon_benchmark_RR2021} and~\ref{fig:gspec_recon_benchmark_RR2021}.

The interpolation tables which comprised the forward models were produced using a Monte Carlo code written in C++. Supplementary note~\ref{sup:RR_models} details the computational implementation of each model of radiation reaction.

\section{Results}\label{sec:results}

A number of difficulties arise when implementing Bayesian inference in a context such as this, where many parameters affect the observables, are poorly constrained and their effects on the observables are correlated. If the latter statement holds, a change in one parameter may be partially compensated for by a change in another parameter. If the prior is insufficiently restrictive, many regions of the parameter space may exist which optimise the posterior: a unique solution may not exist. This problem is known as degeneracy. There is also a risk of overfitting; if the number of inference parameters (i.e. the degrees of freedom of the forward model) are increased, the inference procedure will return progressively larger hyper-volumes of the f-dimensional parameter space which optimise the posterior probability. Without sufficiently constraining priors, additional diagnostics or smaller uncertainties, increasing the number of free parameters may not increase the quantity of meaningful information which can be extracted from the data. Furthermore, if an excessive number of poorly constrained free parameters are used, these parameters may compensate for inaccuracies in the models, allowing any model of radiation reaction to be made compatible with the data. This may be avoided by applying strong priors, however, such priors require measurements of the unknown parameters or must otherwise be physically motivated, and many parameters in collider experiments lack either constraint.

We employ three approaches to address these challenges. To begin with, the collision parameters whose expected variations have the highest and lowest impact on the post-collision electron and gamma spectra are identified. The latter are assigned fixed values in the forward models. Consequently, degeneracies between the parameters which have the greatest impact on the observables are identified, if present. These degeneracies are used to further reduce the number of free parameters; two degenerate parameters may be replaced by one, so long as it is able to accurately reproduce the underlying physics of the collision. 
This inference parameter is then treated as an effective parameter. As such, its inferred value will not reflect a physical property of the electron beam or laser, but rather will reproduce, to first order, the distribution of $\eta$ at the collision. 

Finally, we propose to perform the analysis on shots selected to minimise the potential impact of collision parameters excluded from the forward model. For these shots, degeneracies between included and excluded parameters enable the inference procedure to compensate for any contributions to the electron beam energy loss from parameters not explicitly included in the forward model.

\subsection{Effect of Laser, Electron and Collision Parameters on Post-collision Observables}

\begin{figure}[!ht]
    \centering
    \includegraphics[width=0.6\textwidth, trim={0.0cm 0.0cm 0.0cm 0.0cm}, clip]{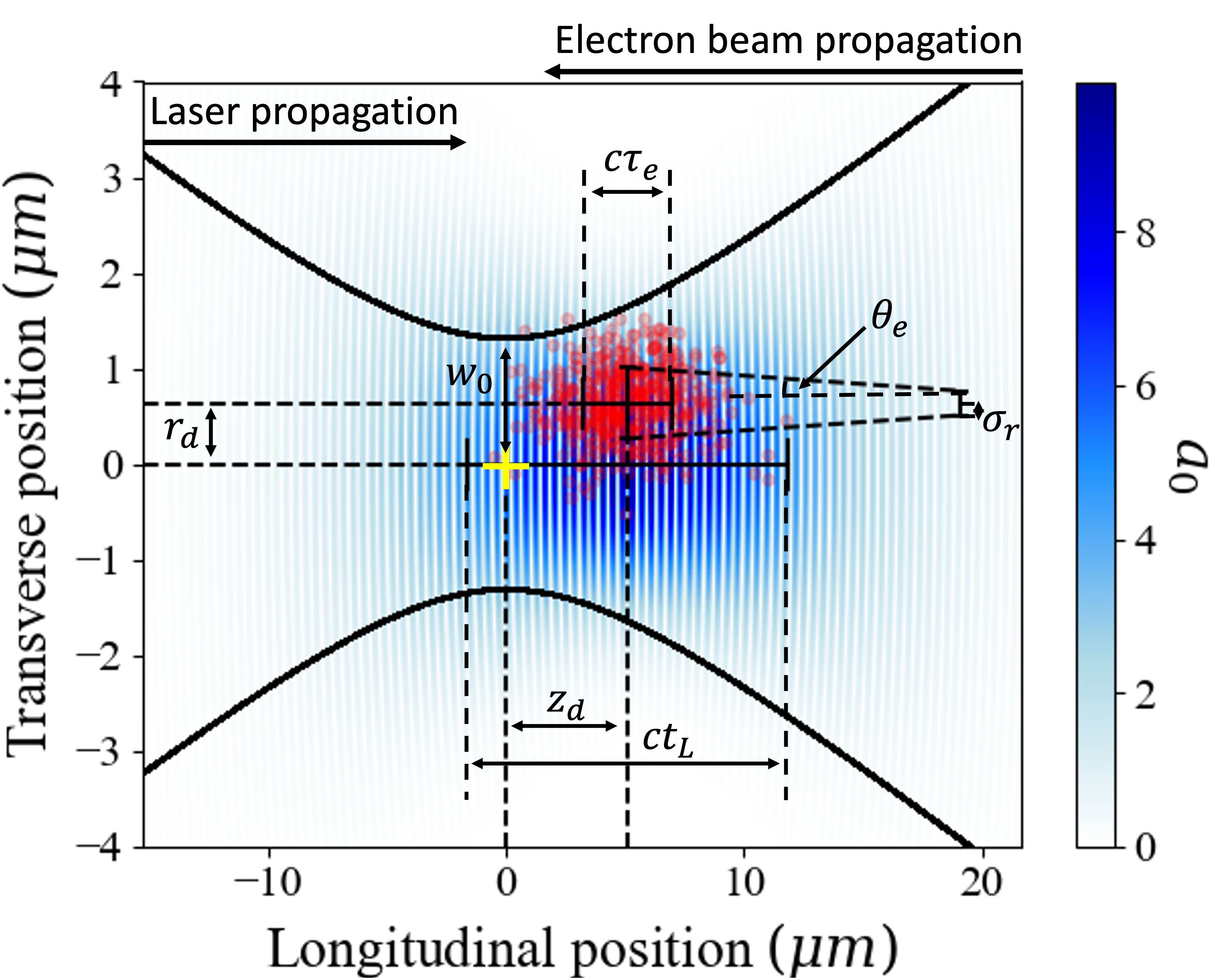}
    \caption{A collision between an electron beam (red) and a tightly focussed, counter-propagating laser (normalised field strength shown in blue) is depicted. The electron beam charge is normally distributed both spatially and temporally, with duration $\tau_e$, source size $\sigma_r$ and energy-dependent divergence $\theta_e$. The laser intensity, which is proportional to the square of the normalised intensity parameter, $a_0$, has gaussian spatial and temporal dependence. The laser waist, $w_0$, and duration, $t_L$, are indicated. The collision is longitudinally and transversely offset from the laser focus (yellow cross) by $z_d$ and $r_d$, respectively. \label{fig:ebeam_laser_collision}}
\end{figure}

The laser, electron and collision parameters which were measured, estimated or inferred from previous measurements are summarised alongside their assigned values in the forward models in tables~\ref{tab:exp_laser_params},~\ref{tab:exp_ebeam_params} and~\ref{tab:coll_params}, respectively. These parameters are illustrated in figure~\ref{fig:ebeam_laser_collision} for clarity. The temporal displacement of the collision from focus provided in table~\ref{tab:coll_params} combines the shot-to-shot variation in the timing between the two laser pulses with the additional delay due to the unknown injection point of the electron beam. As the electron beam is accelerated to velocities exceeding the group velocity of the laser in the plasma, $v_g$, if injection occurs earlier, the colliding beam would need to arrive earlier for the electron beam to interact with the peak laser intensity.

\begin{table}[!ht]
\centering
\begin{tabular}{ccccp{9cm}}
\toprule
Laser parameters &Experiment & Value in forward model\\
\toprule
Energy on target (\SI[]{}{\joule})  &  $6.13\pm 0.02$ & Free parameter\\
FWHM transverse waist (\SI[]{}{\micro\meter\squared}) & $\left(\SI[separate-uncertainty]{2.52(20)}{}\right)\times\left(\SI[separate-uncertainty]{2.09(10)}{}\right)$ & 2.47\\
FWHM duration (\SI[]{}{\femto\second}) & $45\pm3$ & 45\\
\bottomrule
\end{tabular}
\caption{Measured laser parameters.\label{tab:exp_laser_params}}
\end{table}

\begin{table}[!ht]
\centering
\begin{tabular}{cccccp{9cm}}
\toprule
Electron beam property &Experiment&Value in forward model\\ 
\toprule
Duration* (standard deviation) (\SI[]{}{\femto\second}) & \SI[separate-uncertainty]{14(14)}{} & Free parameter\\
Transverse source size& \SI[separate-uncertainty]{0.68(13)}{} & 0.68\\
(standard deviation) (\SI[]{}{\micro\meter})&&\\
Electron beam propagation distance& 0.0&0.0\\
from source to collision plane (\SI[]{}{\milli\meter}) &&\\
Total electron charge (\SI[]{}{\pico\coulomb})& \SI[separate-uncertainty]{140.1(12)}{}& Normalised\\
FWHM divergence (\SI[]{}{\milli\radian}) & $(b_1-b_2\sqrt{\gamma m_e[\SI[]{}{\giga\electronvolt}]})$& $(b_1-b_2\sqrt{\gamma m_e[\SI[]{}{\giga\electronvolt}]})$\\
\bottomrule
\end{tabular}
\caption{Measured or estimated electron beam parameters. simulated parameters are indicated with an asterisk. The electron beam source size has been estimated from previous measurements~\cite{Schnell_2012}, while the electron beam duration was obtained from particle-in-cell simulations using the code FBPIC. The constants $b_1$ and $b_2$ were $b_1=1.30^{+0.26}_{-0.19}~\SI[]{}{\milli\radian}$, $b_2={0.26^{+0.24}_{-0.28}}~\SI{}{\milli\radian\per\giga\electronvolt\tothe{1/2}}$. \label{tab:exp_ebeam_params}}
\end{table}

\begin{table}[!ht]
\centering
\begin{tabular}{ccccp{9cm}}
\toprule
Collision parameters &Experiment& Value in forward model\\
\toprule
Transverse displacement of&  $0.0\pm17.54$ & 0.0\\
collision from focus (\SI[]{}{\micro\meter}) & &\\
Temporal displacement of&  $\pm N(0, 30)$ & Free parameter\\
collision from focus (\SI[]{}{\femto\second}) & $+U(2.73, 45.82)$&
\\
\bottomrule
\end{tabular}
\caption{The expected transverse and temporal alignment of the electron beam and the colliding laser and the expected shot-to-shot jitter in the above parameters. $U$ and $N$ denote uniform and normal distributions, respectively.\label{tab:coll_params}}
\end{table}

\begin{figure}[h]%
\centering
{\begin{overpic}[width=1.0\textwidth]{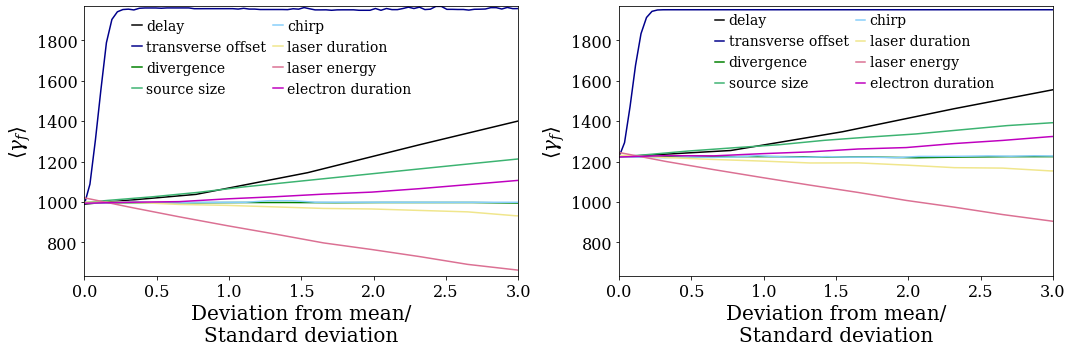}
    \put(23,33){Classical}
    \put(68,33){Quantum-stochastic}
\end{overpic}}
\caption{The mean Lorentz factor of the post-collision electron spectrum predicted by the classical and quantum-stochastic models varies with the deviation of a given collision parameter from its mean value, normalised by the standard deviation.
This choice of normalisation factor illustrates the probability that a parameter will deviate from its mean value by a given amount.\label{fig:hirearchy_coll_params_gammaf}
}
\end{figure}
\begin{figure}[h]%
\centering
{\begin{overpic}[width=1.0\textwidth]{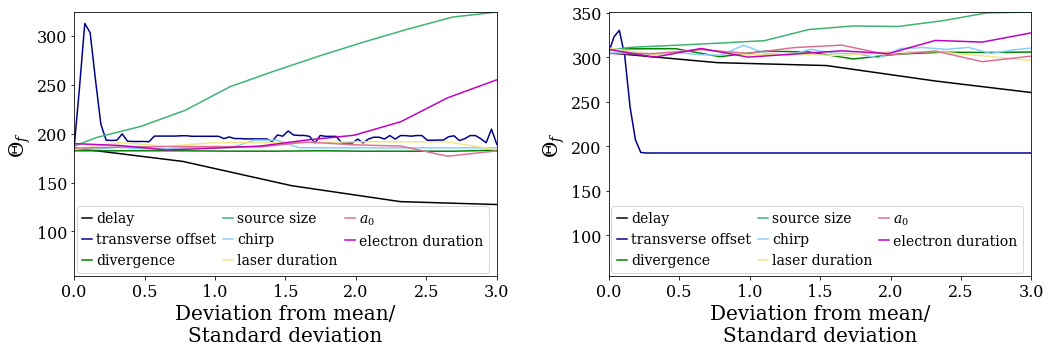}
    \put(23,33){Classical}
    \put(68,33){Quantum-stochastic}
\end{overpic}}
\caption{Similar to figure~\ref{fig:hirearchy_coll_params_gammaf}, where the standard deviation of the post-collision electron spectrum is shown along the y-axis.\label{fig:hirearchy_coll_params_sigmaf}
}
\end{figure}

Figures~\ref{fig:hirearchy_coll_params_gammaf} and~\ref{fig:hirearchy_coll_params_sigmaf} 
illustrate the effect of varying the laser, electron beam and collision parameters on the mean energy and width of the post-collision electron spectrum, respectively. This allows the collision parameters with the highest impact on the experimental observables to be identified. We have chosen to neglect spatio-temporal coupling terms such as pulse front tilt and chirp in the laser, as these higher-order effects are expected to have a significantly lower impact on the post-collision observables compared to the parameters considered here. 
 The variation in the number of emitted photons, $A$, with these parameters has not been shown as this trend is merely the inverse of the trend shown in figure~\ref{fig:hirearchy_coll_params_gammaf}. The variation in the critical energy of the photon spectrum, $\bar{\epsilon}_c$, is not shown as this is much less sensitive to fluctuations in the collision parameters than the properties of the electron spectrum which are shown.


Figure~\ref{fig:hirearchy_coll_params_gammaf} indicates that, given the expected uncertainties in the collision parameters, the transverse offset has the most significant effect upon the electron beam energy loss, followed by the laser energy, the longitudinal offset of the collision from focus and the source size of the electron beam, respectively. By comparison, the effect of changing the electron beam chirp and duration and the laser duration are negligible. The effect of changing the electron beam divergence also appears insignificant, however this is because the longitudinal offset of the collision from the laser focus was set to 0, thus the effect of the changing energy-dependent divergence of the electron beam on its transverse size and (hence the average laser intensity during the collision)  is negligible.

In figure~\ref{fig:hirearchy_coll_params_sigmaf}, the expected transverse jitter also has the highest expected impact on the electron spectrum, however the most impactful parameters following this are the electron source size, the longitudinal offset and the electron beam duration.

The laser $a_0$, longitudinal offset of the collision from the laser focus, $Z_d$, and the electron beam duration, $\tau_e$ were chosen to be free parameters in the inference procedure. As the laser duration and waist at focus remained constant, a change in laser $a_0$ indicates a change in the laser energy. The laser $a_0$ was chosen as its expected effect on the electron energy losses was larger than any other parameter, barring the transverse jitter. This also allowed an arbitrary number of shots with differing laser energies to be analysed using the same forward model. The longitudinal offset was selected due to the significance of its impact on both the mean and width of the post-collision electron spectrum. Finally, the electron beam duration was chosen over the source size, which has a greater effect on the width of the post-collision electron spectrum, as degeneracy allowed the effect of changing source size to be compensated for by varying $Z_d$, as will be discussed in section~\ref{sec:degeneracy}. Finally, only the shots with the highest photon yields were analysed to reduced the probability of analysing a collision transversely offset from the laser focus. For this reason, the transverse offset in the forward models was fixed at 0. This is discussed in greater detail in section~\ref{sec:degeneracy}. 

\subsection{Degeneracy}\label{sec:degeneracy}

\begin{figure}[h]%
\centering
\includegraphics[width=0.55\textwidth]{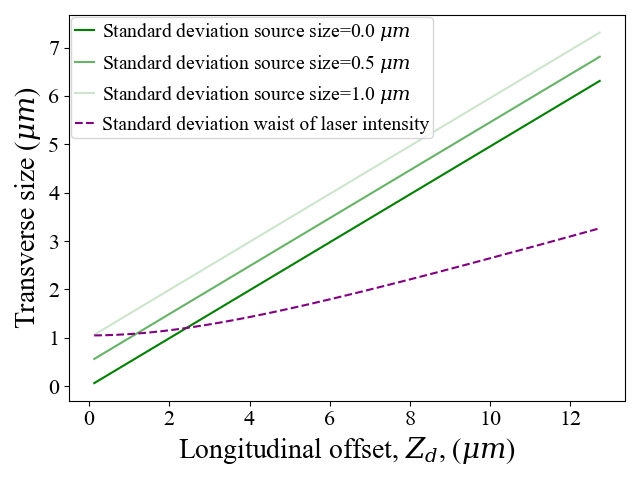}
\caption{The effect of electron beam divergence and source size on the relative transverse sizes of the electron beam and colliding laser is shown as a function of longitudinal displacement from the electron beam source and the laser focus, respectively.\label{fig:source_size_div_degeneracy}
}
\end{figure}

\begin{figure}[h]%
\centering
{\begin{overpic}[width=1.0\textwidth]{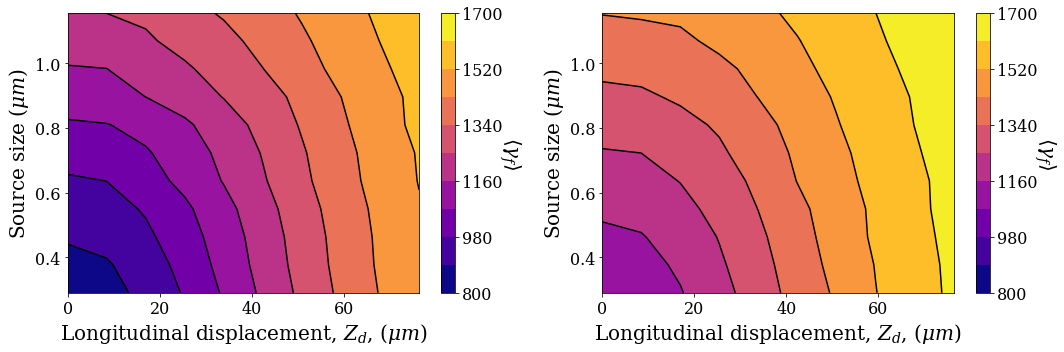}
    \put(20,33){Classical}
    \put(63,33){Quantum-stochastic}
\end{overpic}}
\caption{The location of the post-collision electron Lorentz factor, $\langle\gamma_f\rangle$, as a function of electron beam source size and longitudinal displacement of the collision from the laser focus for the classical and quantum-stochastic models. 
\label{fig:delay_vs_ss_gammaf}
}
\end{figure}

\begin{figure}[h]%
\centering
{\begin{overpic}[width=1.0\textwidth]{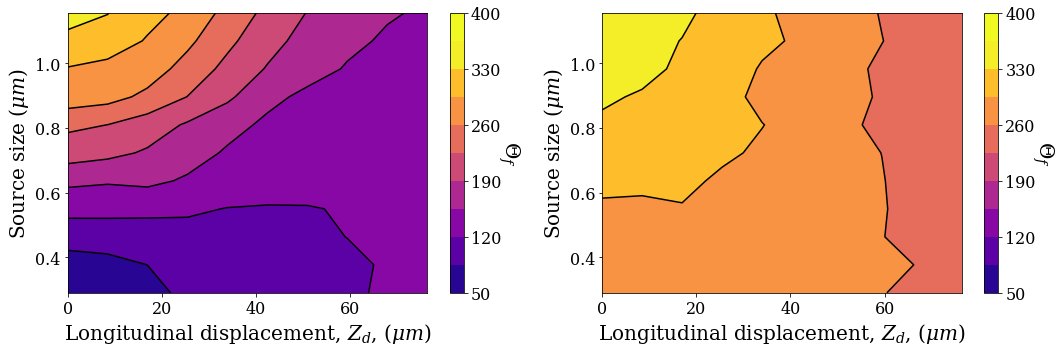}
    \put(20,33){Classical}
    \put(63,33){Quantum-stochastic} 
\end{overpic}}
\caption{The scale of the electron spectrum, $\Theta_f$, predicted by the classical and quantum-stochastic models of radiation reaction as the electron beam source size and the longitudinal displacement of the collision from the laser focus are varied.\label{fig:delay_vs_ss_sigmaf}
}
\end{figure}

A change in source size alters the relative transverse sizes of the laser and electron beam. However, as illustrated in figure~\ref{fig:source_size_div_degeneracy}, the energy-dependent electron beam divergence, which is included in the forward models as a fixed parameter, produces a similar effect for a given longitudinal displacement. This indicates that the effect of a changing source size may be compensated for if $a_0$ and $Z_d$ are free parameters. This is further supported by figures~\ref{fig:delay_vs_ss_gammaf} and~\ref{fig:delay_vs_ss_sigmaf}, which indicate that degeneracies exist between $Z_d$ and initial source size for both $\langle \gamma_f \rangle$ and $\Theta_f$.

The transverse jitter is by far the most impactful collision parameter. The combination of the large shot-to-shot variation in the electron beam pointing and the steep radial dependence of the intensity of the colliding laser ensure that the electron beam energy loss becomes negligible if the transverse offset is even $0.5\sigma$ from perfect alignment for $Z_d=0$. As the transverse offset was expected to affect the electron beam energy loss more severely than any other parameter, both in terms of the magnitude of its effect on the electron beam energy loss and the high probability of a transverse misalignment due to large shot-to-shot variations in the electron beam pointing, the shots which produced the highest gamma yield out of a dataset of successful collisions were more likely to be well-aligned transversely than longitudinally.
For this reason, we propose that the Bayesian analysis should be applied to shots which produce the highest photon yields. This increases the probability that the transverse offset is small for the selected shots, and can thus be compensated for by exploiting the degeneracy between the laser $a_0$, $Z_d$ and $\tau_e$ and the transverse offset.%
\begin{figure}[h]%
\centering
{\begin{overpic}[width=1.0\textwidth]{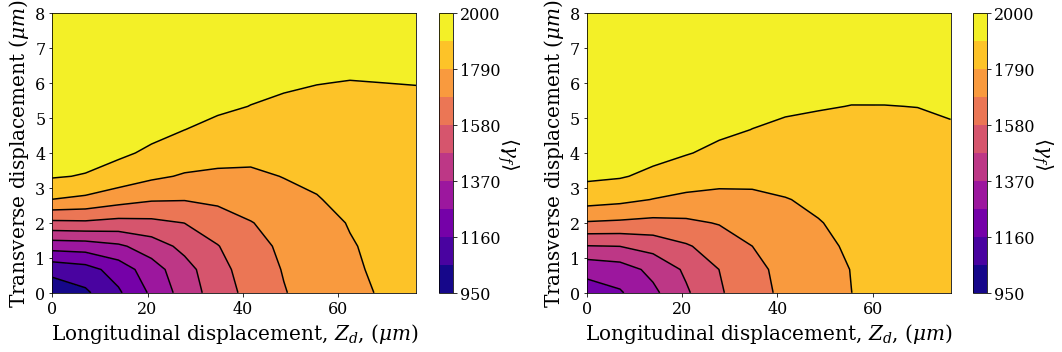}
    \put(19,33){Classical}
    \put(62,33){Quantum-stochastic}
\end{overpic}}
\caption{The location, $\langle \gamma_f \rangle$, of the post-collision electron Lorentz factor distribution predicted by the classical and quantum-stochastic models of radiation reaction is shown with varying longitudinal and transverse displacement of the collision from the laser focus.\label{fig:delay_vs_transverse_offset_gammaf}
}
\end{figure}

\begin{figure}[h]%
\centering
{\begin{overpic}[width=1.0\textwidth]{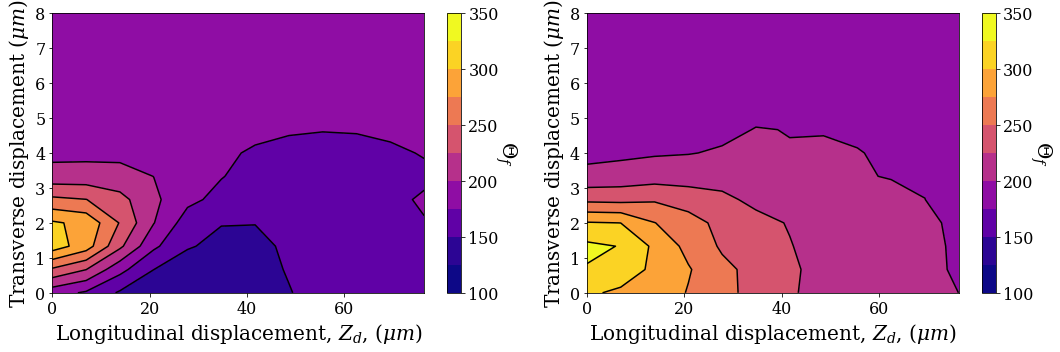}
    \put(19,33){Classical}
    \put(62,33){Quantum-stochastic}
\end{overpic}}
\caption{The scale of the post-collision electron Lorentz factor distribution, $\Theta_f$, predicted by the classical and quantum-stochastic models of radiation reaction for varying transverse and longitudinal alignment between the electron beam and the colliding laser.\label{fig:delay_vs_transverse_offset_sigmaf}
}
\end{figure}

Figures~\ref{fig:delay_vs_transverse_offset_gammaf} and~\ref{fig:delay_vs_transverse_offset_sigmaf} demonstrate the existence of degeneracies between $Z_d$ and transverse offset, as was the case for source size and $Z_d$. However, while degeneracies are evident in $\langle \gamma_f \rangle$ and $\Theta_f$ when these observables are considered separately, for the inference framework to exploit the degeneracies between free and fixed parameters, it must identify a combination of $a_0$, $Z_d$ and $\tau_e$ which reproduces these observables simultaneously.

\subsection{Bayesian Test Cases}

\begin{figure}[!p]
    \centering

     \begin{subfigure}[t]{0.85\textwidth}
      \begin{minipage}[t]{0.55\textwidth}
        \centering
        \vspace{-1.0em}
        \captionsetup{labelformat=empty}
        \caption{Classical}
        \label{subfig:cl_error_with_displ_from_ideal_params}
    \end{minipage}
    \hfill 
    \begin{minipage}[t]{0.49\textwidth}
        \centering
        \vspace{-1.0em}
        \captionsetup{labelformat=empty}
        \caption{Stochastic}
        \label{subfig:st_error_with_displ_from_ideal_params}
    \end{minipage}
    \includegraphics[width=0.99\textwidth, trim={0.1cm 0.2cm 0.1cm 0.1cm}, clip]{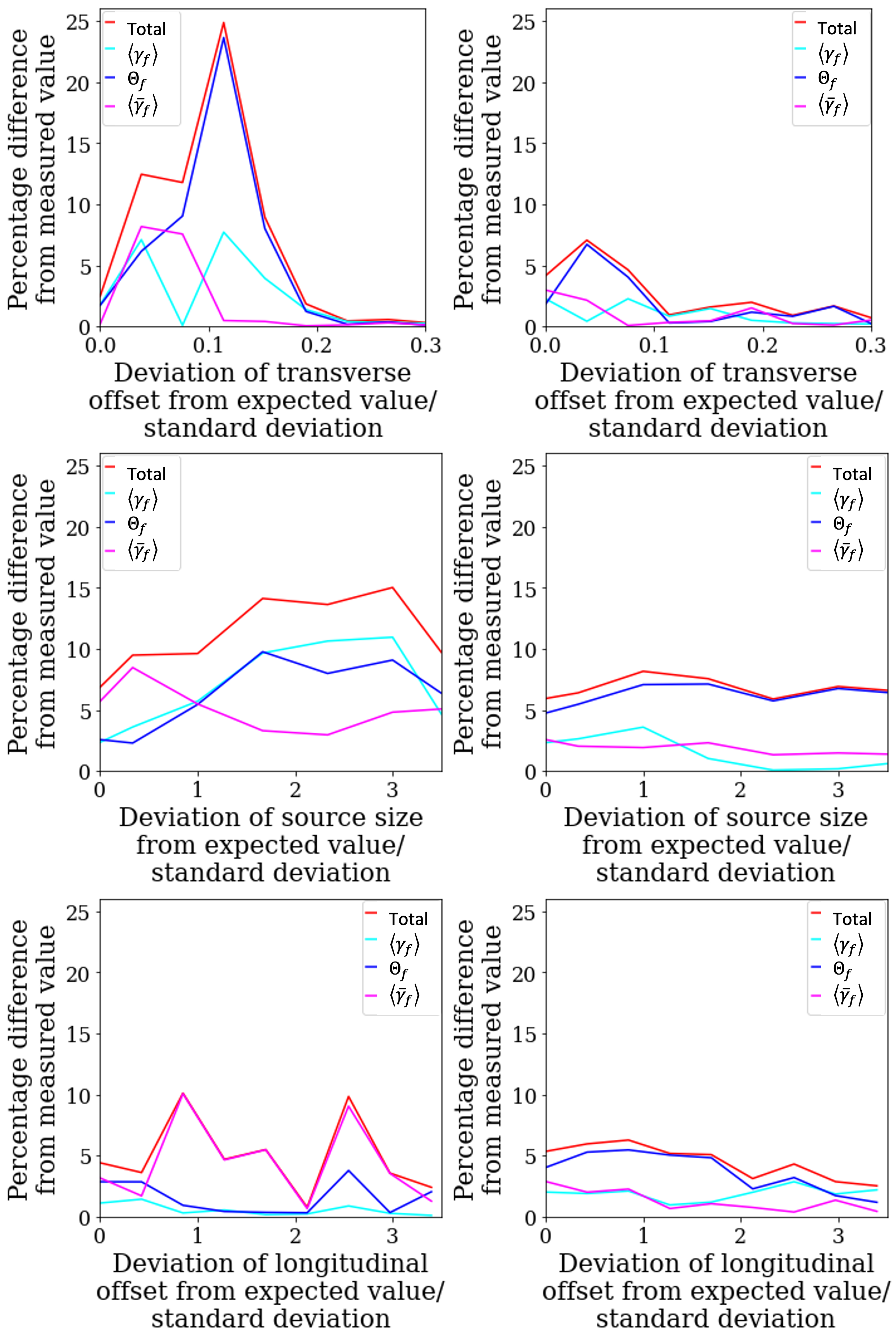}
    \end{subfigure}
    \vspace{-1.0em}
    \caption{The percentage difference between the simulated and inferred values for the average, $\langle\gamma_f \rangle$, and standard deviation, $\Theta_f$ of the post-collision electron Lorentz factor distribution, and the average energy of the photon distribution, $\langle\bar{\epsilon}_f \rangle$, are shown as the longitudinal and transverse offset of the collision from the laser focus and the electron beam source size are varied. The total error is given by the root mean squared deviation of the inferred $\langle\gamma_f \rangle$, $\Theta_f$ and $\langle\bar{\epsilon}_f \rangle$ from the simulated values.\label{fig:gammaf_sigmaf_transverse_offset}}
\end{figure}

We investigated whether the Bayesian inference procedure treats free parameters as effective parameters, i.e., uses free parameters to reproduce the effects on post-collision observables of fixed parameters which differ from their set values in the forward models. To this end, the procedure was performed on a series of simulated post-collision electron and gamma spectra for each model of radiation reaction. In these simulations, either the longitudinal offset, transverse offset or the electron beam source size was varied. Each inference procedure then fitted the simulated data using the corresponding forward model (i.e. the data simulated using the classical model was fitted using the classical inference procedure). For each inference procedure, the parameters fixed in the forward models had the same values, given in tables~\ref{tab:exp_laser_params},~\ref{tab:exp_ebeam_params} and~\ref{tab:coll_params}. The inference results are summarised in figure~\ref{fig:gammaf_sigmaf_transverse_offset}.

Degeneracies between electron beam duration and longitudinal offset of the collision from focus allow the quantum-stochastic model to recover the post-collision electron and gamma spectra with a total error (which combines the error in the inferred mean, standard deviation of the electron spectrum mean of the gamma spectrum) which is typically at the few percent level, and is always $<10\%$.
The maximum total error in the classical model is $25\%$, however for the majority of the parameter values considered in figure~\ref{fig:gammaf_sigmaf_transverse_offset}, the uncertainty $<10\%$.
For both models, the uncertainty in the inferred spectra for varying source size (a fixed parameter) is only a few percent higher than the inferences performed for varying longitudinal offset (a fitted parameter), confirming the high degree of degeneracy between these parameters and bolstering our decision to include the latter but not the former. 
The effect of the transverse offset is harder to replicate: this motivates our decision to down-select our data to reduce the probability that a large transverse offset was present for the collisions we analysed using the Bayesian framework. As the gamma radiation yield scales as $\propto \tilde{a}_0 Q\gamma^2$~\cite{poder_2018}, where $\tilde{a}_0$ is the effective laser $a_0$ and $Q$ is the total electron beam charge, by selecting shots with the high gamma yield normalised by $Q\gamma^2$, the probability that the collision is transversely offset from the laser focus for the selected shots is minimised.

A set of inference procedures were run for mono-energetic electron beams with mean energy $\SI[]{1}{\giga\electronvolt}$, in which the transverse offset in the simulated collision was progressively increased, for the following collision parameters; $a_0=21.38$, $Z_d=0$, $\tau_e=\SI[]{20}{\femto\second}$. It was found that model differentiation was no longer possible for a transverse offset of $1.5w_0=\SI[]{3.3}{\micro\metre}$, at which the reduction in the effective collision $a_0$ and the spectral broadening induced by the transverse offset rendered model selection infeasible.

Three further test cases were performed to determine whether the inference procedure is able to extract the correct (input) model of radiation reaction and the correct collision parameters for electron spectra and uncertainties representative of experimental conditions, as these spectra are broadband and have complex, non-normal charge distributions. To achieve this, the post-collision observables were simulated for collisions characterised by different parameters in each test case. Three inferences were then performed on the data produced by each simulation; one per model.

\begin{figure}[!ht]
    \centering
    \begin{subfigure}[t]{0.99\textwidth}
    	{\begin{overpic}[width=0.99\textwidth, trim={0.0cm  0.1cm 0.2cm 0.2cm}, clip]{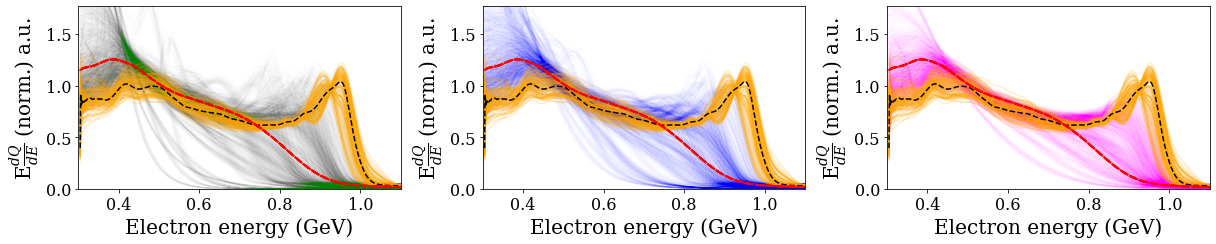}
     \put(-2,18){a)}
\put(14,20.5){Classical}
    \put(40.5,20.5){Quantum-continuous}
	\put(74.5, 20.5){Quantum-stochastic}
\end{overpic}}
   \end{subfigure}
    \vspace{0.2em}
    \\
     \begin{subfigure}[t]{0.99\textwidth}
    {\begin{overpic}[width=0.99\textwidth, trim={0.0cm 0.2cm 0.2cm 0.0cm}, clip]{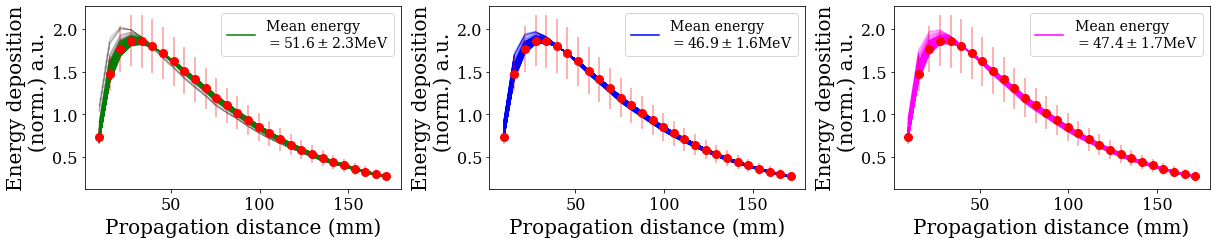}
    \put(-2,18){b)}
    \end{overpic}}
    \end{subfigure}
    \caption{The quantum-stochastic model of radiation reaction was used to simulate a collision between a focussing, gaussian laser pulse with $a_0=16$, $Z_d=\SI[]{30}{\femto\second}$ and $\tau_e=\SI[]{14}{\femto\second}$ (the remaining laser and electron beam parameters are provided in tables~\ref{tab:exp_laser_params} and~\ref{tab:exp_ebeam_params}, respectively) and the pre-collision electron spectrum. Simulated data and classical, quantum-continuous and quantum-stochastic inferences are shown in red, green, blue and magenta, respectively. This colour scheme will be used consistently for the remaining figures in this section. a) The simulated post-collision electron spectrum, predicted pre-collision electron spectra (orange), and its median (black), alongside the inferred post-collision electron spectra. b) The simulated and inferred responses of the photon spectrometer as a function of photon propagation depth.\label{fig:ideal_espec}}
\end{figure}

For the first test case, described as ``ideal'', the stochastic model was used to simulate the collision and the simulation parameters have identical values to their fixed counterparts in the forward models. This means that in both the simulation and the forward model, the transverse offset of the collision from the laser focus is zero, the electron beam source size is \SI[]{0.68}{\micro\meter}, etc. This test case probes the ability of the inference procedure to retrieve the collision parameters and perform model comparison accurately for an ideal scenario, where the collision conditions which are fully described by the forward model. The inferred post-collision electron spectra and photon spectrometer responses for this test case are compared to the simulated data in figure~\ref{fig:ideal_espec}.

\begin{figure}[h]%
\centering
{\begin{overpic}[width=1.0\textwidth]{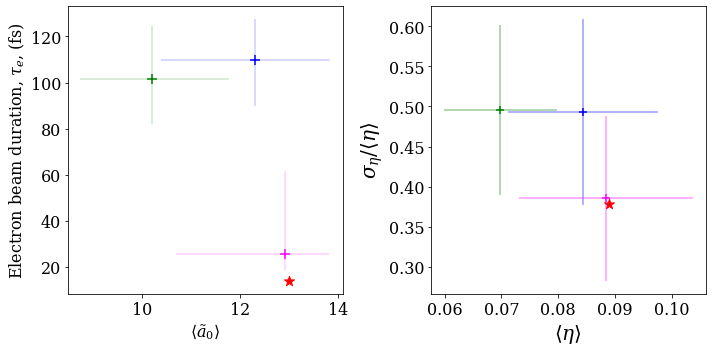}
 \put(1,48){a)}
\put(51, 48){b)}
\end{overpic}}
\caption{Inference parameters obtained for the first test case, where the stochastic model was used to simulate the collision.
The collision parameters inferred by the classical (green), quantum-continuous (blue) and quantum-stochastic (magenta) models compared to the simulation input parameters (red star). a) $\langle \tilde{a}_0\rangle$, the average effective collision $a_0$ the electron beam interacts with during the collision. The collision distribution of $\langle \tilde{a}_0\rangle$ stems from the finite size of the electron beam, the spatio-temporal dependence of laser intensity, and their overlap. Hence $\langle \tilde{a}_0\rangle$ is a function of all three inference parameters. b) The mean and standard deviation of the collision distribution of $\eta$ due to the broadband electron spectrum and the range of $\tilde{a}_0$ the electron beam experiences during the collision.\label{fig:fake_espec}}
\end{figure}

In figure~\ref{fig:fake_espec}, the quantum-stochastic and quantum-continuous models retrieve $\langle \tilde{a}_0\rangle$, $\langle\eta \rangle$ and $\sigma_\eta$ within $1\sigma$ of the input parameters. None of the models infer $\tau_e$ within $1\sigma$ of the input value. For the classical and quantum-continuous models, this indicates that $\tau_e$ has been treated as an effective parameter. Rather than reflecting the true electron beam duration, the inferred value of $\tau_e$ allows the distribution of effective $a_0$ and $\eta$ to be most closely reproduced. Of the three models, the quantum-stochastic model infers the value of $\tau_e$ most accurately, but fails to recover the correct value with $1\sigma$, indicating that the uncertainty in the pre-collision electron spectrum and the lack of sensitivity of the post-collision electron spectrum to small variations in $\tau_e$ inhibit accurate inferences of this parameter. The classical and quantum-continuous inferences both retrieve $\tau_e$ considerably greater than the true value, thereby increasing the range of $\tilde{a}_0$ with which the electron beam interacts and hence the range of energy losses it experiences. This allows the classical and quantum-continuous models to reproduce the spectral broadening induced by the quantum-stochasticity inherent in the quantum-stochastic model, which the classical and quantum-continuous models do not predict. As the classical model predicts higher energy losses than the quantum-stochastic model for equivalent collision parameters, it infers a lower $\langle \tilde{a}_0\rangle$ to produce comparable energy losses. 

The $\langle \tilde{a}_0\rangle$ and standard deviation $\tilde{a}_0$ (the latter results from the spatial and temporal overlap of the electron beam and laser) are reflected in the mean, $\langle\eta\rangle$ and standard deviation, $\sigma_{\eta}$, of the distribution of $\eta$ which characterises the collision.

When comparing the quantum-stochastic and classical models, $r_{\rm{qs}, \rm{cl}}=1.2$, while the quantum-continuous and classical models yield $r_{\rm{qc}, \rm{cl}}=0.9$, and the quantum-stochastic and quantum-continuous model comparison gives $r_{\rm{qs}, \rm{qc}}=1.3$, indicating there is insufficient evidence to differentiate between the quantum-stochastic, quantum-continuous and classical models for the simulated collision conditions, given the uncertainties on the predicted pre-collision electron spectrum and the measured post-collision electron and gamma spectra. 

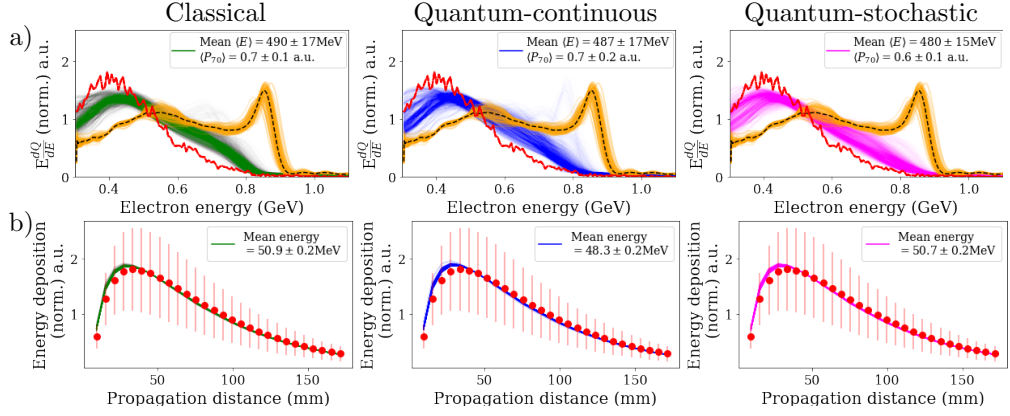
\begin{figure}[!ht]
    \centering
    \begin{subfigure}[t]{0.99\textwidth}
    	{\begin{overpic}[width=0.99\textwidth, trim={0.2cm  0.1cm 0.2cm 0.2cm}, clip]{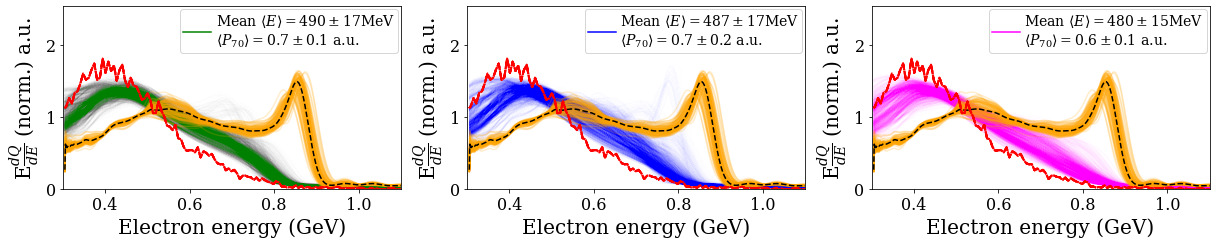}
     \put(-2,18){a)}
\put(14,20.5){Classical}
    \put(39.5,20.5){Quantum-continuous}
	\put(73.5, 20.5){Quantum-stochastic}
 \end{overpic}}
   \end{subfigure}
    \vspace{0.0em}
    \\
     \begin{subfigure}[t]{0.99\textwidth}
    {\begin{overpic}[width=0.99\textwidth, trim={0.2cm 0.3cm 0.2cm 0.2cm}, clip]{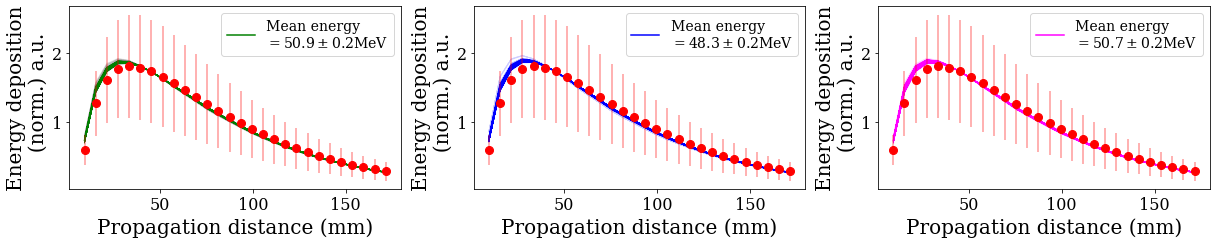}
    \put(-2.5,18){b)}
    \end{overpic}}
    \end{subfigure}
    \caption{The classical model of radiation reaction was used to simulate a collision between a focussing, gaussian laser pulse with $a_0=21.38$, $Z_d=\SI[]{30}{\femto\second}$, $\tau_e=\SI[]{14}{\femto\second}$ and and electron beam, which were offset transversely by \SI[]{1.05}{\micro\metre} (the remaining laser and electron beam parameters are provided in tables~\ref{tab:exp_laser_params} and~\ref{tab:exp_ebeam_params}, respectively). a) The simulated post-collision electron spectrum, predicted pre-collision electron spectra (orange), and its median (black), alongside the inferred post-collision electron spectra. b) The simulated and inferred responses of the photon spectrometer as a function of photon propagation depth.\label{fig:trans_cl_espec}}
\end{figure}

For the second test case, the results of which are shown in figure~\ref{fig:trans_cl_espec}, the classical model was used to simulate a collision which was transversely offset from the laser propagation axis by \SI[]{1.05}{\micro\metre}. Three inference procedures, one for each radiation reaction model, were performed on the simulated electron and photon spectra. 
The presence of a finite transverse offset between the electron beam and laser focus at the collision induces electron spectral broadening, which resembles the spectral broadening predicted by the quantum-stochastic model. This test case illustrates the extent to which the inference procedure is able to perform model selection accurately (i.e. select the classical model), if the collision parameters with fixed values in the forward models differ from those values. This is particularly pertinent if the classical model is accurate and the additional collision parameters induce spectral broadening, an effect also predicted by the quantum-stochastic model. This test case also indicates whether and how inference procedures use degeneracy to compensate for collision parameters which differ from the fixed values in the forward model.

\begin{figure}[h]%
\centering
{\begin{overpic}[width=1.0\textwidth]{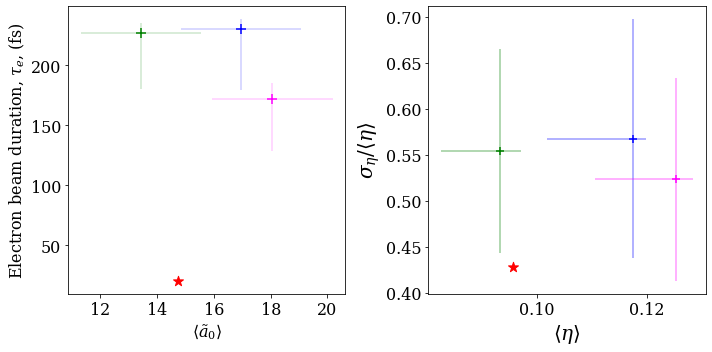}
 \put(1,48){a)}
\put(51, 48){b)}
\end{overpic}}
\caption{Similar to figure~\ref{fig:fake_espec}, where the input and inferred parameters pertain to the transversely offset classical test case.\label{fig:trans_cl_params}}
\end{figure}

The inferred and input collision parameters for the second test case in figure~\ref{fig:trans_cl_params} indicate that only the classical model is able to infer the mean effective $a_0$ of the collision and the mean $\eta$ within $1\sigma$. However, all models, including the classical model, over-estimate the electron beam duration and $\sigma_\eta$. This contrasts with the first test case in which the correct model (i.e. the quantum-stochastic model) also inferred $\sigma_{\eta}$ within $1\sigma$ of its true value. In the second test case, all models treat $\tau_e$ as an effective parameter, stretching the electron beam longitudinally to replicate (in so far as possible) the broad distribution of $a_0$ with which the electron beam interacted due to the finite transverse offset. This is also reflected in the overly large $\sigma_{\eta}$ inferred.
For the second test case (see figure~\ref{fig:trans_cl_espec}), each of the models appear to infer the post-collision electron and gamma spectra with comparable accuracy. This is substantiated by the Bayes factors; we obtain $r_{\rm{qs}, \rm{cl}}=1.0$, while $r_{\rm{qc}, \rm{cl}}=0.7$, and $r_{\rm{qs}, \rm{qc}}=1.4$, thus there is insufficient evidence to favour any of the models.


In the third test case, the collision was simulated using the quantum-stochastic model and was transversely offset by \SI[]{3.2}{\micro\metre} from the laser propagation axis. This allowed the model selection capability of the Bayesian analysis to be verified for a collision with parameters which differed from their fixed values in the forward model, in which the quantum-stochastic model was used to produce the test data. The results for the third test case are provided in supplementary note~\ref{sup:test_case3_results}. For the third test case, the Bayes factor obtained when comparing the quantum-stochastic and classical models,
$r_{\rm{qs}, \rm{cl}}=1.8$, the quantum-continuous and classical models yield $r_{\rm{qc}, \rm{cl}}=1.7$ while a comparison of the quantum-stochastic and quantum-continuous inferences yields $r_{\rm{qc}, \rm{qs}}=1.0$, thus there is insufficient evidence to differentiate between the quantum-stochastic, quantum-continuous and classical models.

Each test case yielded $0.33<r_{\rm{cl}, \rm{qs}}<3.2$, indicating model selection could not be performed for a single shot given the collision conditions and uncertainties considered for each test case. However, for both test cases in which the quantum-stochastic model was used to generate the test data, $r_{\rm{cl}, \rm{qs}}>1.0$, indicating model selection would be feasible if evidence were combined across multiple shots. For the ideal test, assuming model evidence is consistent for each shot, seven shots would be required to allow model differentiation.

For the test case in which the classical model was used to generate the test data, $r_{\rm{qs}, \rm{cl}}=1.0$ and $r_{\rm{qc}, \rm{cl}}<1.0$, indicating that if the collision parameters induce spectral broadening, the stochastic model would not be unduly favoured over the classical model.

We have ascertained that the quantum-stochastic model retrieves $\langle \tilde{a}_0\rangle$, $\langle\eta\rangle$ and $\sigma_{\eta}$ within $1\sigma$ of the input parameters when this model is used to generate the simulated test data and the forward model accurately describes the full complexity of the collision. As illustrated in the classical test case, if additional collision parameters which are not present in the forward model are present in the simulated data, the correct model (i.e. the classical model) infers $\langle \tilde{a}_0\rangle$ and $\langle\eta\rangle$ within $1\sigma$ of the simulation values, but significantly overestimates $\tau_e$ and hence $\sigma_\eta$. This indicates that while the distributions of $\eta$ and $\langle \tilde{a}_0\rangle$ are inferred accurately to first order, there are insufficient degrees of freedom in the models to accurately infer higher order moments (i.e $\sigma_{\tilde{a}_0}$, $\sigma_\eta$) if collision parameters which are fixed in the forward models differ significantly from these values.

\section{Conclusion}

We have developed a novel Bayesian framework which infers values of unknown collision parameters and predicts corresponding experimental observables for the classical, quantum-continuous and quantum-stochastic models of radiation reaction. We identify challenges associated with the application of a Bayesian approach to this problem, such as overfitting and insufficiently constraining priors. We address these issues by down-selecting the number of free parameters and the data to be analysed and by exploiting degeneracies between free and fixed parameters. This has motivated the choice of $a_0$, $z_d$ and $\tau_e$ as fitting parameters and the decision to fix the remaining parameters.

We demonstrate that the Bayesian framework consistently infers $\langle \tilde{a}_0\rangle$, $\langle\eta \rangle$, $\sigma_\eta$ accurately (within $1\sigma$) for the highest-performing model. Quantitative comparisons of the relative performance of these models for a single shot yield insubstantial evidence in favour of the correct model over other models, indicating that while model discrimination may not be feasible at the single shot level, this may be accomplished by combining evidence across multiple shots. We find that when the transverse offset differs from its fixed value in the forward models, the standard deviations of the distributions of $\langle \tilde{a}_0\rangle$ and $\eta$ are inferred less accurately, however accurate model selection with this framework is still robust for transverse offsets up to \SI[]{3.3}{\micro\metre} for the electron beam and laser parameters considered.

The inclusion of free parameters such as the transverse offset would be facilitated if the computational expense (and runtime) of the Bayesian analysis were reduced. This could be accomplished if the electron spectra were mono-energetic and if strong priors could be applied to restrict the available parameter space and avoid overfitting. Additionally, improved laser stability would facilitate a greater fraction of collisions with good spatial-temporal overlap between the electron beam and colliding laser, increasing the number of collisions for which quantum effects are expected to be substantial. 

We anticipate that Bayesian inference will prove to be a powerful analysis tool for the interpretation of future strong field QED experiments involving colliding lasers and particle beams, and have demonstrated the feasibility and utility of such an analysis for an all-optical radiation reaction experiment.

\appendix
\renewcommand{\thesection}{\Alph{section}}
\renewcommand{\thesubsection}{\Alph{section}.\arabic{subsection}}

\section{Additional results}\label{sup:test_case3_results}

\begin{figure}[!ht]
    \centering
    \begin{subfigure}[t]{0.99\textwidth}
    	{\begin{overpic}[width=0.99\textwidth, trim={0.2cm  0.1cm 0.1cm 0.1cm}, clip]{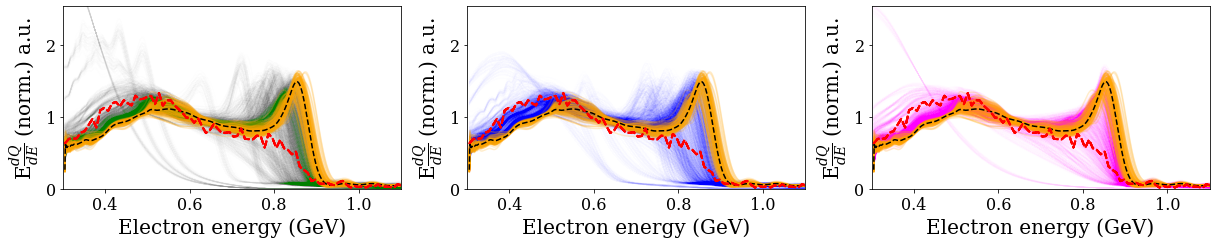}
     \put(-2,18){a)}
    \put(14,20.5){Classical}
    \put(39.5,20.5){Quantum-continuous}
	\put(73.5, 20.5){Quantum-stochastic}
    \end{overpic}}
   \end{subfigure}
    \vspace{0.0em}
    \\
     \begin{subfigure}[t]{0.99\textwidth}
    {\begin{overpic}[width=0.99\textwidth, trim={0.2cm 0.3cm 0.1cm 0.1cm}, clip]{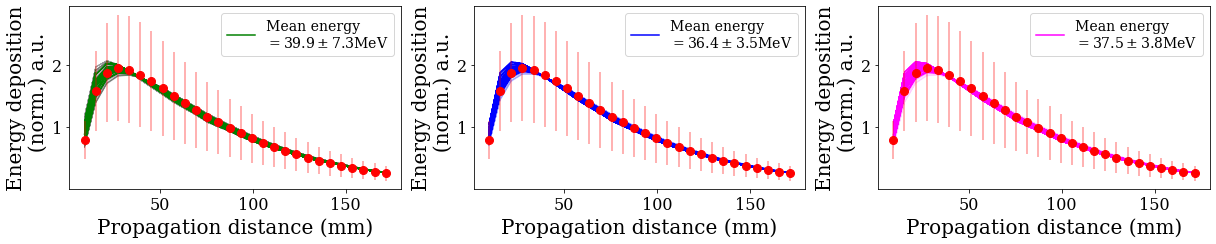}
    \put(-2.5,18){b)}
    \end{overpic}}
    \end{subfigure}
    \caption{The quantum-stochastic model of radiation reaction was used to simulate a collision between a focussing, gaussian laser pulse with $a_0=21.38$, $Z_d=\SI[]{30}{\femto\second}$, $\tau_e=\SI[]{20}{\femto\second}$ and the transverse offset is \SI[]{2.1}{\micro\metre} (the remaining laser and electron beam parameters are provided in tables~\ref{tab:exp_laser_params} and~\ref{tab:exp_ebeam_params}, respectively) and the pre-collision electron spectrum. Simulated data and classical, quantum-continuous and quantum-stochastic inferences are shown in red, green, blue and magenta, respectively. a) The simulated post-collision electron spectrum, predicted pre-collision electron spectra (orange), and its median (black), alongside the inferred post-collision electron spectra. b) The simulated and inferred responses of the photon spectrometer as a function of propagation depth.\label{fig:trans_st_espec}}
\end{figure}

\begin{figure}[h]%
\centering
{\begin{overpic}[width=1.0\textwidth]{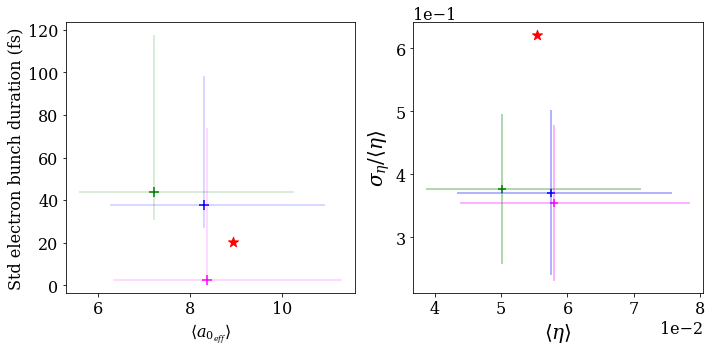}
 \put(1,48){a)}
\put(51, 48){b)}
\end{overpic}}
\caption{Similar to figure~\ref{fig:fake_espec}, where the input and inferred parameters pertain to the transversely offset stochastic test case.\label{fig:trans_params}}
\end{figure}

\section{Supplementary Note}\label{sup:RR_models}

A quantum treatment of radiation reaction in a strong field requires a direct calculation of the scattering matrix for an electron interacting with an arbitrary electromagnetic field, as perturbation theory is invalid in this regime. For fields with spatial and temporal structure, such calculations are not tractable. To surmount this difficulty, the Furry picture is utilised, in which the interaction of an electron with a classical background field is integrated into its basis state. Perturbation theory may then be applied to these states to describe quantum photon scattering processes. We also use the ``locally-constant field approximation'' or LCFA~\cite{Nikishov_1964}, expected to be valid for $a_0\gg1$ and $\frac{a_0^3}{\eta}\gg1$~\cite{Niel_2018}. The photon emission rates derived under the LCFA are used to formulate the equations of motion for an electron interacting with a strong external field. Between photon emissions, electron motion is classical. The equations of motion are propagated in time using a fourth-order Runge-Kutta algorithm in C++~\cite{Kirk_2009}\cite{Arber_2015}\cite{Duclous_2010}.

The classical theory of radiation reaction predicts that the reaction force manifests as a higher-order correction term in the relativistic equation for the Lorentz four-force~\cite{LL_1971_ClassicalTheoryFields}
\begin{equation}
\label{4_vec_eom_mov_charge_w_rad_reaction_cl}
m_e \frac{du^\alpha}{ds}=eF^{\alpha\beta}u_\beta-P_0\frac{u^\alpha}{c^2}.
\end{equation}
where $P_0=\frac{m_ec^5e^2}{6\pi\epsilon_0\hbar}$, $m_e$, $e$ and $u_\alpha$ are the electron mass, charge and four-velocity, respectively, $F^{\alpha\beta}$ is the electromagnetic tensor, $c$ is the speed of light in vacuum, $\hbar$ is the reduced Plank's constant, $\epsilon_0$ is the permittivity of free space and $\eta$, the electron quantum parameter, is defined in section~\ref{sec:intro}. 

Photon emission is accounted for as follows. The classical rate of photon emission, $\dot{\bar{N}}^{cl}$,~\cite{Blackburn_2020} 
\begin{equation}
\label{eqn:cl_opt_depth}
\dot{\bar{N}}^{cl}=\frac{5\alpha m_ec^2 \chi}{2\sqrt{3}\hbar\gamma}.
\end{equation}
is integrated over time, $t$. Here, the quantum photon parameter, $\chi\approx\frac{\bar{\epsilon} b}{2}$, $b=\frac{1}{\alpha E_{sch}}\sqrt{(\vec{E}+\vec{v}\times\vec{B})^2-(\vec{E}\cdot\vec{v}/c)^2}$, $\vec{E}$ and $\vec{B}$ are the electric and magnetic fields, respectively, $\vec{v}$ the electron velocity vector and $\alpha$ is the fine structure constant. A photon is emitted when $T^{cl}\geq-\ln(1-\Lambda^{cl})$, where $\Lambda^{cl}$ is sampled randomly from a uniform distribution, $U[0, 1]$~\cite{Arber_2015}. 

The polar angle, $\theta$, at which the photon was emitted was sampled from the differential rate of photon emission $\frac{\partial\dot{\bar{N}}}{\partial z}$ with respect to $z=\left[2\gamma^2(1-\beta\cos\theta)\right]^{3/2}$~\cite{Blackburn_2020}
\begin{equation}
\label{eqn:cl_d2N_dt_dz}
\frac{\partial\dot{\bar{N}}^{cl}}{\partial z}=\frac{\alpha m_ec^2\chi}{\sqrt{3}\hbar\gamma}\frac{z^\frac{2}{3}-\frac{1}{2}}{z^2}
\end{equation}
where $\beta=|\vec{v}|/c$. This closely resembles the approach employed by Duclous et. al.~\cite{Duclous_2010} and Arber et. al.~\cite{Arber_2015} where the photon energy is obtained by sampling from the differential rate of emission with respect to energy.

Given $z$, the photon energy is computed from the differential probability of photon emission, $\frac{\partial^2\dot{\bar{N}}}{\partial u\partial z}$,~\cite{Blackburn_2020}
\begin{equation}
\label{eqn:cl_d3N_dZ_du_dt}
\frac{\partial^2\dot{\bar{N}}^{cl}}{\partial u\partial z}=\frac{2\alpha m_ec^2}{3\sqrt{3}\hbar\pi\chi\gamma}u(2z^\frac{2}{3}-1)R_\frac{1}{3}\left(\frac{2uz}{3\chi}\right).
\end{equation}
where $u=\frac{\bar{\epsilon}}{\gamma - \bar{\epsilon}}$, $\bar{\epsilon}=\frac{\hbar\omega}{m_ec^2}$ and $R_\frac{1}{3}$ is a modified Bessel function of the second kind~\cite{Ternov_1995}. The azimuthal angle, $\phi$, is sampled from a uniform distribution, $U[0,2\pi]$.

Under the quantum-continuous model, the reaction force term in equation~\ref{4_vec_eom_mov_charge_w_rad_reaction_cl} is modified by the Gaunt factor, $g(\eta)$~\cite{Baier_1998},
\begin{equation}
\label{4_vec_eom_mov_charge_w_rad_reaction_sc}
m_e \frac{du^i}{ds}=eF^{ik}u_k-g(\eta)P_0\frac{u^i}{c^2}.
\end{equation}
where $g(\eta)$ is defined as~\cite{Niel_2018}\cite{Kirk_2009}
\begin{equation}
\label{g_eta_def}
g(\eta)=\frac{9\sqrt{3}}{8\pi}\int^\infty_0 dy\left[\frac{2y^2R_{5/3}(y)}{(2+3\eta y)^2}+\frac{36\eta^2y^3K_{2/3}(y)}{(2+3\eta y)^4}\right]
\end{equation}
and is well-approximated by~\cite{Ridgers_2017}
\begin{equation}
\label{g_eta_approx}
g(\eta)\approx \left[1+4.8(1+\eta)\ln(1+1.7\eta)+2.44\eta^2)\right]^{-2/3}.
\end{equation}

The rate of photon emission is sampled from the quantum differential rate, $\dot{\bar{N}}^{st}$,~\cite{Blackburn_2020}\cite{Kirk_2009}
\begin{equation}
\label{st_optical_depth_def}
\dot{\bar{N}}^{st}=\frac{\alpha m_ec^2}{3\sqrt{3}\pi\hbar\gamma}\int^\infty_0\frac{5u^2+7u+5}{(1+u)^3}K_{2/3}du
\end{equation}
where $K_{2/3}$ is a Bessel function of the second kind.

The photon energy is sampled from the quantum differential rate, $\frac{d\dot{\bar{N}}}{d\bar{\epsilon}}$~\cite{Niel_2018}
\begin{equation}
\label{quant_number_spec_gamma_gamma}
\frac{d\dot{\bar{N}}^{st}}{d\bar{\epsilon}}=\frac{\alpha m_ec^2}{\sqrt{3}\hbar\pi\gamma^2}\left\{\left[\frac{2}{3\eta y}+\frac{3\eta y}{2}\right]K_{2/3}(y)-\int^\infty_yK_{1/3}(s)ds\right\}
\end{equation}

Given $\bar{\epsilon}$, the polar angle at which the photon is emitted is sampled from~\cite{Blackburn_2020}
\begin{equation}
\label{quant_number_spec_angular}
\frac{\partial^3\dot{\bar{N}}^{st}}{\partial u\partial z\partial \phi}=\frac{\alpha m_ec^2}{3\sqrt{3}\hbar \pi^2 \eta_s\gamma}\frac{u}{(1+u)^3}\left[z^{2/3}(2+2u+u^2)-(1+u)\right]K_{1/3}\left(\frac{2uz}{3\eta_s}\right).
\end{equation}
As with the classical model, $\phi$ is sampled from a uniform distribution, $U[0,2\pi]$.

The equation of motion for an electron under the quantum-stochastic model of radiation reaction in between emission events is merely the Lorentz force with no additional reaction term~\cite{Duclous_2010}
\begin{equation}
\label{4_vec_eom_mov_charge_w_rad_reaction_st}
m_e \frac{du^i}{ds}=eF^{ik}u_k
\end{equation}

As in the quantum-continuous model, the rate of photon emission is sampled from equation~\ref{st_optical_depth_def}, the photon energy and polar angle are sampled from equations~\ref{quant_number_spec_gamma_gamma} and~\ref{quant_number_spec_angular}, respectively, and $\phi$ is sampled from a uniform distribution, $U[0,2\pi]$. However, unlike the quantum-continuous approach the effect of the recoil on the electron energy and trajectory is then calculated using equation~\ref{eqn:st_electron_e_loss}~\cite{Duclous_2010}:
\begin{equation}
\label{eqn:st_electron_e_loss}
p'^{\alpha}=p^{\alpha}-q^{\alpha}
\end{equation}
where $p^{\alpha}$ and $p'^{\alpha}$ are electron four-momenta before and after the emission, and $q^{\alpha}$ is the photon four-momentum.

\bibliographystyle{unsrt}
\bibliography{methodsRR2021}

\end{document}